\documentclass[epj]{svjour}
\usepackage{graphicx,float,amsmath,amsfonts,color,unitsdef}
\usepackage[sort&compress,numbers]{natbib}

\newcommand{\eps}{\varepsilon}
\begin{document}

\title{Droplets on liquids and their long way into equilibrium}
\author{Stefan Bommer\inst{1} \and Sebastian Jachalski\inst{2} \and Dirk Peschka\inst{2} \and Ralf Seemann \inst{1,3} \and  Barbara Wagner \inst{2,4}}

\offprints{}          
\institute{Experimental Physics, Saarland University, D-66123 Saarbr\"ucken, Germany \and
Weierstrass Institute, Mohrenstr.~39, D-10117 Berlin, Germany \and
Max Planck Institute for Dynamics and Self-Organization, D-37073 G\"ottingen, Germany \and
Institute of Mathematics, Technische Universit\"at Berlin, Stra{\ss}e des 17.~Juni 136, D-10623 Berlin, Germany}

\date{Received: date / Revised version: date}
%

\abstract{The morphological paths towards equilibrium droplets during the late
stages of the dewetting process of a liquid film from a liquid substrate is
investigated experimentally and theoretically. As liquids, short chained
poly\-styrene (PS) and polymethyl-methacrylate (PMMA) are used, which can be
considered as Newontian liquids well above their glass transition temperatures.
Careful imaging of the PS/air interface of the droplets during equilibration by
\emph{in situ} scanning force microscopy and the PS/PMMA interface after removal
of the PS droplets reveal a surprisingly deep penetration of the PS droplets
into the PMMA layer. Droplets of sufficiently small volumes develop the typical
lens shape and were used to extract the ratio of the PS/air and PS/PMMA surface
tensions and the contact angles by comparison to theoretical exact equilibrium
solutions of the liquid/liquid system. Using these results in our dynamical
thin-film model we find that before the droplets reach their equilibrium they
undergo several intermediate stages each with a well-defined signature in shape.
Moreover, the intermediate droplet shapes are independent of the details of the
initial configuration, while the time scale they are reached depend strongly on
the droplet volume. This is shown by the numerical solutions of the thin-film
model and demonstrated by quantitative comparison to experimental results.}

%
\maketitle

\section{Introduction}

Understanding the dynamics and evolution of patterns in micro- and nanoscale
two-layer immiscible thin films is of fundamental importance for the design of
functional surfaces for numerous applications ranging from model for tear film
of the human eye \cite{ZMC03} to instabilities during the fabrication of
polymer/polymer solar cells \cite{HJ05}. The past decades has seen many
investigations of such systems, both experimentally and theoretically. But
besides of its technological relevance and its scientific interest, the in-depth
understanding of these processes is still superficial compared to liquid
dewetting from solid substrates and quantitative experimental verification is
still missing. For experimental reasons the morphology and the dynamics of
liquid/liquid dewetting was typically studied using polymers as high molecular
liquids. For the model two-layer system of  poly\-styrene (PS) on
polymethyl-methacrylate (PMMA), the shape of an underlying liquid (PMMA) and the
liquid poly\-styrene (PS) rim profile dewetting from this substrate has been
studied experimentally in the pioneering work of the group of G. Krausch 
\cite{lambooy1996dewetting,wang2001dewetting}. As a result, they found a
characteristic shape and dewetting dynamics, depending on the relative viscosity
of the two liquids. The experimentally observed behavior was claimed to be in
agreement with another important work by Brochard-Wyart et al.
\cite{brochard1993liquid}. However, this is surprising since the dewetting
velocity strongly depends on film thickness and which are not considered in
\cite{brochard1993liquid}. However, the used polymers in
\cite{lambooy1996dewetting,wang2001dewetting} are above the entanglement of the
respective chain length and viscoelastic properties cannot be ruled out in this
system. Other groups studied the breakup and the hole growth of a liquid-liquid
system where the viscosity of one of the liquids is much larger than the
viscosity of the other liquid  \cite{segalman1999dynamics} and in the special
case, where the resulting dewetting morphologies are coated with a thin layer of
the underlying liquid \cite{slep00}, whereas the characteristic shape of the
liquid-liquid interface was not explored in detail. Similar results on dewetting
rim shapes can be found also in 
\cite{pan1997unstable,segalman1999dynamics,li2005surface,neto2006novel,
HSJJSDW02,de2007switching} for different polymer-polymer dewetting systems.
However, in none of those experimental studies the parameters are worked out
well enough to allow for a quantitative comparison with theoretical results.
While many of the theoretical works have focussed on linear stability analysis
and numerical simulations of short time and long-time evolution
\cite{pototsky2005morphology,fisher2005nonlinear}, Golovin and Fisher
\cite{fisher2005nonlinear,bandyopadhyay2005instability} even in the presence of
surfactant \cite{fisher2007instability}, a systematic comparison to experimental
results for the simplest possible system will be important to build a
fundamental understanding of this system.

In this study we present a quantitative comparison of experimentally obtained
transient and equilibrium drop shapes to numerically and analytically drop
shapes for a model system using short-chained PS/PMMA, that can be described via
a Stokes model for the multiphase flow of two immiscible liquids with capillary
boundaries and driven by the intermolecular forces between the liquids. In our
experiments we obtain 3d dynamic and equilibrium droplets shapes by following
the complete dewetting process. The relevant parameters required for the
comparison with theoretical derived dynamic droplet shapes, \textit{i.e.} the surface
tension ratio and the contact angle, are obtained by comparing the experimental
equilibrium droplet shapes to the exactly know theoretical shapes. These values
are used as input parameters in our dynamic thin-film model to compare the
various morphological transitions long before reaching equilibrium. Except for a
short initial phase the morphology was found to be independent on the initial
start conditions. This independence of the transient droplet shapes allows for
quantitative comparison of the experimentally determined drop morphologies to
the theoretically droplet shapes calculated using the experimentally determined
input parameters.

\section{Methods}

For our liquid/liquid experiments we prepared thin poly\-styrene (PS) films on top of underlying polymethyl-meth\-acrylate (PMMA) films which are supported by silicon wafers: Silicon rectangles of about $2\squarecentimeter$ were cut from 5''-wafers with
$\langle 100 \rangle$ orientation. The pieces were pre-cleaned by a fast CO$_\textrm{2}$-stream (snow-jet, Tectra) and sonicated in ethanol, acetone and toluene followed by a bath in peroxymonosulfuric acid and a careful rinse with Millipore$^\textrm{TM}$ water. On top
of the cleaned silicon wafers, PMMA films were spun from toluene solution with thicknesses ranging from $80\nm$ to $700\nm$.  The dewetting PS films cannot be spun directly on top of the PMMA with the desired thickness and were, in a first step, spun from toluene solution onto freshly cleaved mica sheets with thicknesses ranging from $7\nm$ to $22\nm$. In a second step, the glassy PS films were transferred onto a water surface (Millipore$^\textrm{TM}$) and picked up with the PMMA coated silicon substrates.

Both polymers were purchased from Polymer Standard Service Mainz (PSS-Mainz,Germany) with $M_w/M_n=1.03$ and molecular weights of $M_w=9.6\kilogram/\mole$ for PS and $M_w=9.9\kilogram/\mole$ in case PMMA. The molecular weights are well below the entanglement lengths of PS and PMMA and their melts can be treated as Newtonian liquids  \cite{seemann2001shape,becker2003complex,seemann2001dewetting,pechhold,fet2007} at temperatures above their bulk glass transition temperatures $T_G$ of about $T_\textrm{G,PS}=85 \pm 5\celsius$ and $T_\textrm{G,PMMA}=115\pm 5\celsius$ for PS and PMMA, respectively.
The glassy sample can be stored in the lab at room temperature and the liquid/liquid dewetting process is started when heating the sample above the glass transition temperature of the polymers. The dewetting experiments were typically conducted at a temperature of $T=140\celsius$ resulting in viscosities of $\mu_\textrm{ PS}\approx 2\,\kilo\pascal\second$ for the dewetting PS and $\mu_\textrm{PMMA}=675\,\kilo\pascal\second$ for the underlying PMMA film. The viscosity values were extracted from \cite{herminghaus2001glass,baumchen2012slippage}.

\begin{figure}
\centering
\includegraphics[width=0.49\textwidth]{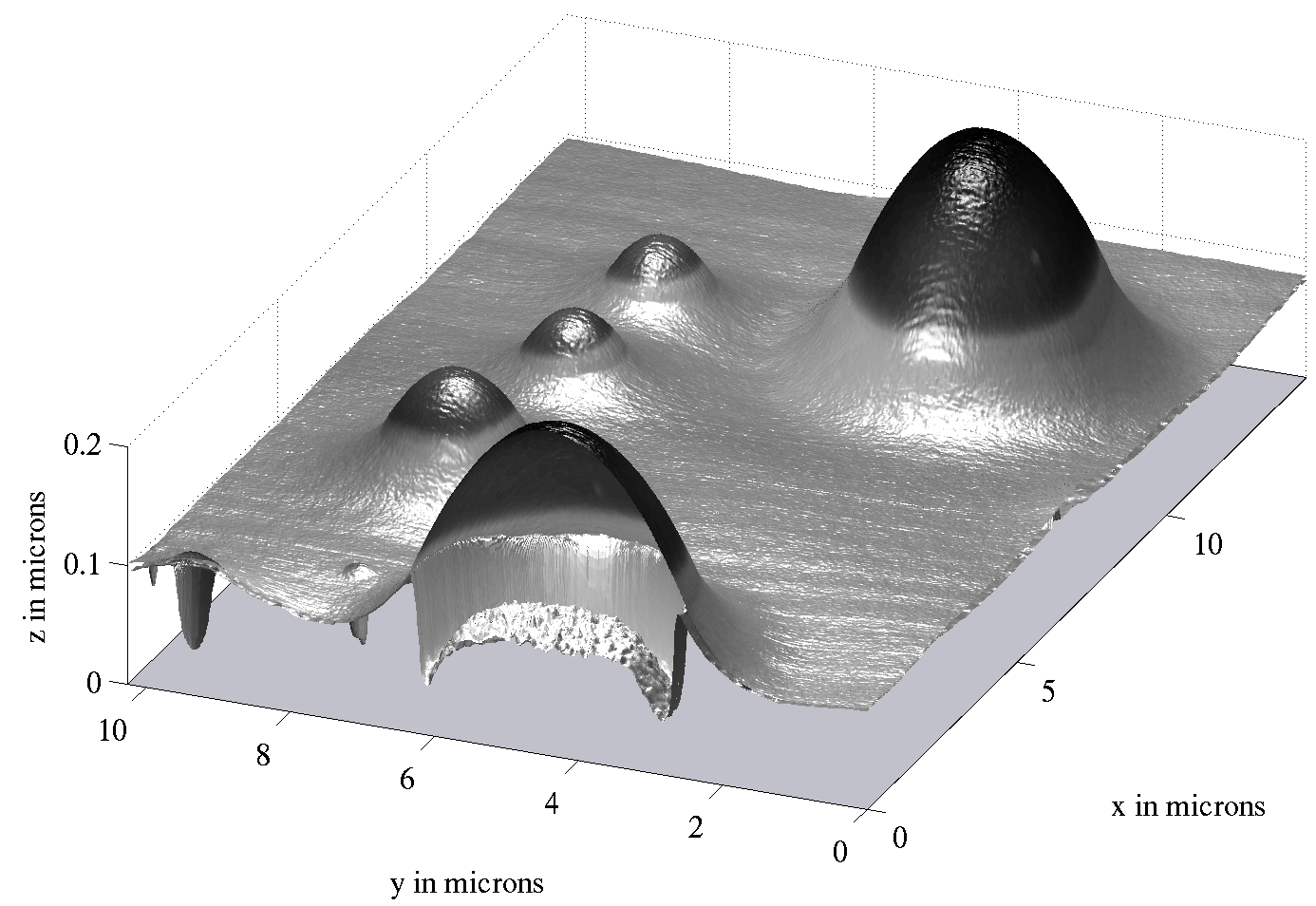}
\caption{Composition of AFM top and bottom topographies after dewetting at $T=140\celsius$
for $45\minute$. Initial top layer thickness $80\nm$ PMMA and bottom layer thickness
$20\nm$ PS layer lead to isolated droplets, which are not yet in equilibrium}
\label{fig:droplet_present}
\end{figure}

The dewetting process was monitored \emph{in situ} at $140\celsius$ by atomic force microscopy (AFM) in Fastscan Mode$^\textrm{TM}$ (Bruker, Germany). To determine additionally the shape of the liquid PS/PMMA interface the dewetting process is stopped at a desired dewetting stage by quenching the sample temperature from $T\,=\,140\celsius$ to room temperature. At this temperature both polymers are glassy and the sample can be easily stored and handled. The glassy PS structures are removed by a selective solvent (cyclohexane,
Sigma Aldrich, Germany) and the remaining PMMA surface with the shape of the formerly PS/PMMA interface frozen into it is subsequently imaged by AFM. The full three dimensional shape of the dewetting PS structures are obtained by combining the surfaces of the related topographies as determined by AFM as shown in fig.~\ref{fig:droplet_present}.

\section{Results and discussion}
\subsection{Equilibrium droplets}

In this section we aim to characterize the subsequently used Si/PMMA/PS/air system by its equilibrium configuration, \textit{i.e.} liquid PS lenses swimming on liquid PMMA supported by silicon. We will first discuss the experimental method to measure droplet shapes, give explicit expressions for equilibrium states and describe how surface tension values can be extracted from those and finally compute contact angles.

In previous experiments no change of the dewetting profile could be detected when cooling and freezing a poly\-styrene dewetting structure on a solid substrate, \textit{e.g.} \cite{seemann2001shape}.  However, in our liquid/liquid system it was found that the PS/air interfaces significantly change their shape upon cooling. A PS droplet that raised out of the PMMA by \textit{e.g.} 12\nm at $T\,=\,140\celsius$ just raises out of the PMMA by 1-2\nm at room temperature.  And as we aim at a quantitative comparison with theoretical results it is of utmost importance to clarify the origin of the shape change and to develop a suitable experimental protocol.

\begin{figure}
\centering
\includegraphics[height=3.5cm]{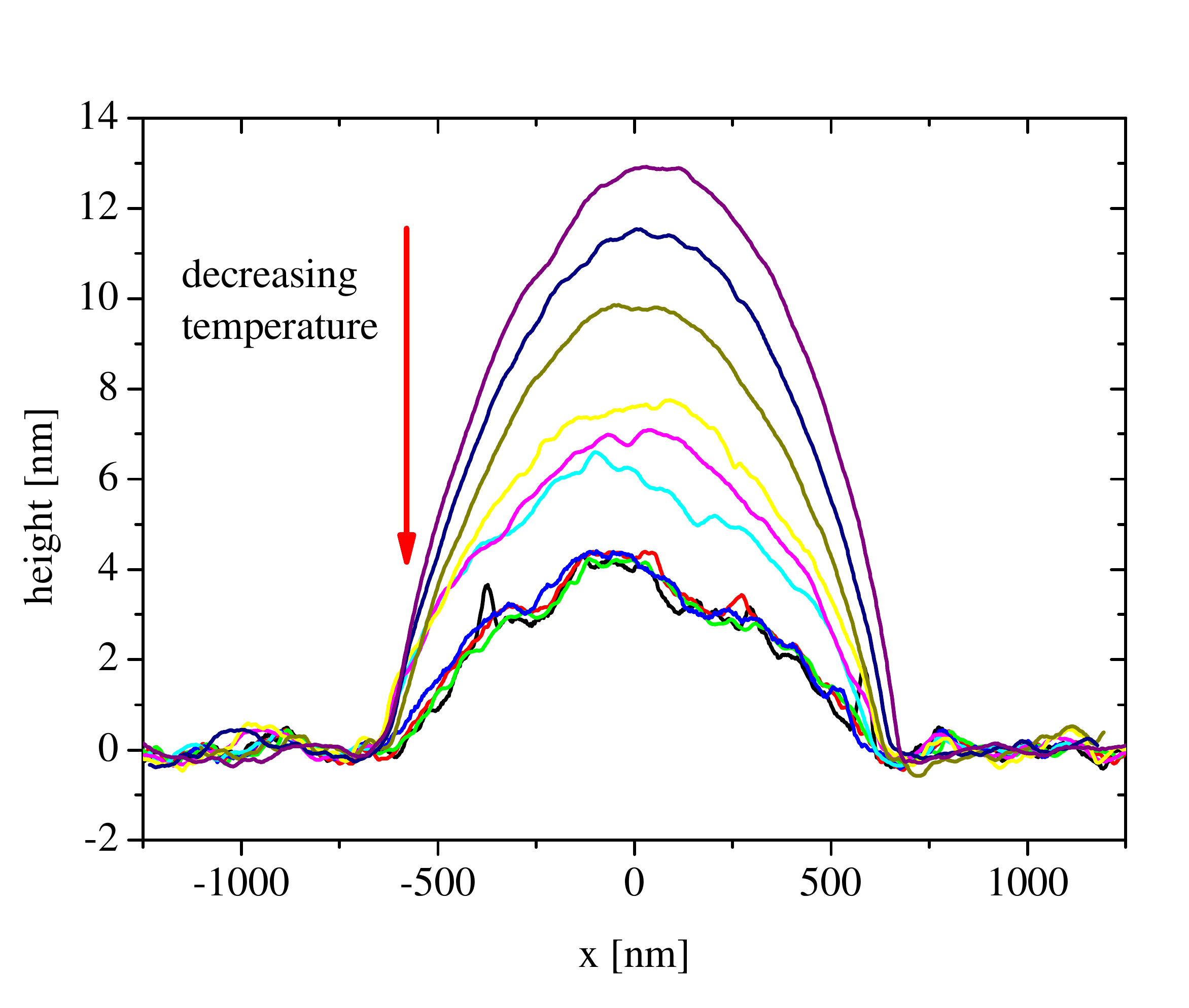}%
\includegraphics[height=3.5cm]{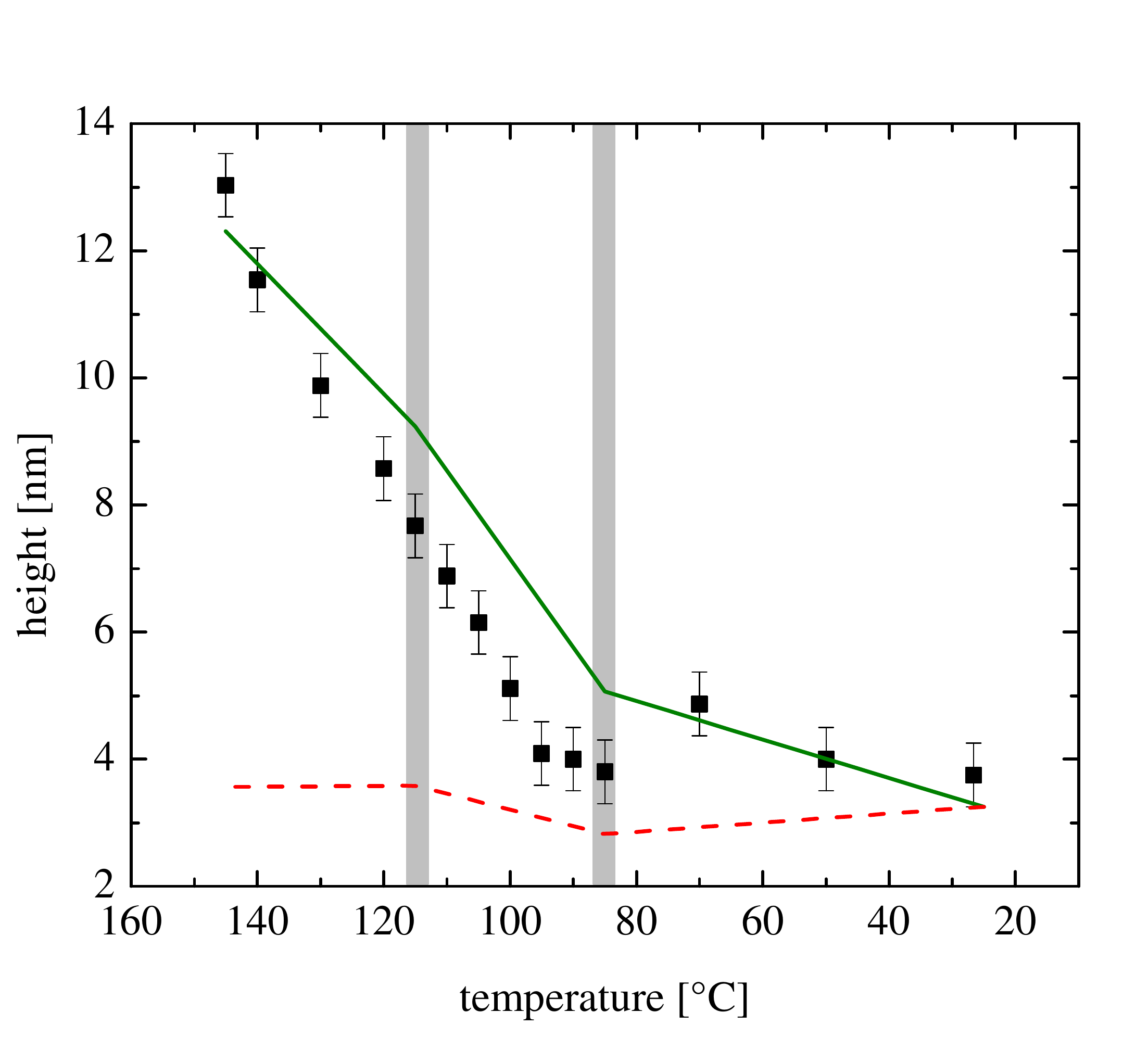}
\caption{(left) Cross sections of the same poly\-styrene droplet imaged \emph{in situ} by AFM
 while cooling the sample from $145\celsius$ to $30\celsius$. The entire duration of this cooling
 cycle was about 70\,minutes. (right) Temperature dependent height of the droplet as obtained from AFM data
 with glass transition temperatures of PS and PMMA are indicated. The solid line shows the expected
 height change based on thermal expansion
 polymers when assuming a constant width of the droplet. The dashed line denotes a similar model
 calculating assuming just the linear expansion coefficients.}
\label{fig:freeze1}
\end{figure}

To observe the shape change of the PS/air interface during a temperature change, the PS/air interfaces of several droplets were measured while cooling and heating. As a first result, the shape change during heating and cooling cycles turned out to be reversible, see fig.~\ref{fig:freeze1} for a cooling cycle. Considering the viscosities of both polymers as well as the experimentally found independence of the droplet shape from the cooling/heating rate (tested from milliseconds to hours), a lateral equilibration
\textit{e.g.} according to temperature dependent surface tensions is not expected. This could be confirmed experimentally as no dependence of the in-plane radius of the droplet  with temperature could be detected within an experimental error of about $\pm 10\nm$. However, the height of the droplet changes significantly showing a characteristic step in the shape change as function of temperature, see fig.~\ref{fig:freeze1}

Even not knowing the thermal expansion coefficients precisely, as slightly
different values can be found in literature, the height change of the droplet
can be explained by simple model assumptions. The approximate thermal  volume
expansion coefficients of PS are $\alpha_{\rm PS,T < T_G}\approx 2 \cdot
10^{-4}$, $\alpha_{\rm PS,T > T_G}\approx 5.5 \cdot 10^{-4}$ and for PMMA
$\alpha_{\rm PMMA,T < T_G}\, \approx \, 2.7 \cdot 10^{-4}$, $\alpha_{\rm PMMA,T
> T_G}=5.4 \cdot 10^{-4}$ \cite{krevelen1976}. The remarkably change of
expansion coefficient at the glass transition temperature is characteristic for
polymers:  At room temperature both polymers are glassy whereas the $E$
module of  the PMMA is larger than that of PS. Accordingly the shape of the
PS/PMMA interface is expected to remain constant during the temperature increase
and the entire thermal volume  expansion of the PS is used to increase the
height of the droplet. Reaching the  glass transition temperature of PS, the PS
expands almost three times faster with temperature, \textit{i.e.} twice as fast as the
still glassy PMMA. This change in thermal expansion coefficient thus explains
the pronounced step in the droplet height for temperatures $T > T_\textrm{G,PS}$.
When reaching the glass transition temperature of PMMA, the expansion ratios of
both polymers are again similar. Still assuming that no lateral equilibration of
 the lower droplet shape occurs, the height change of droplet with temperature
is just slightly reduced as we still expect the difference between the linear
expansion for PMMA and the volume expansion of PS, see solid line in
fig.~\ref{fig:freeze1}\,(right panel). Repeating the same reasoning assuming an
isotropic change in droplet volume we have to calculate the height change of the
droplet based on the linear expansion coefficients of both polymers which are
three times smaller than the corresponding volume expansion coefficients. The
result of this model calculation is plotted as dashed line in
fig.~\ref{fig:freeze1}\,(right panel). Comparing the result of both model
calculations to the experimental data we can conclude that the height change of
the droplets results in fact from the thermal expansion of the polymers  while
the PS/PMMA interface remains about constant during this temperature induced
shape change.

Based on this result we validated our experimental protocol to obtain
quantitative drop shapes: The top side of the droplets, \textit{i.e.} the PS/air
interface is imaged at the applied dewetting temperature and combined with the
PMMA/PS interface measured at room temperature after removing the PS structure.
\begin{figure}
\centering
\includegraphics[height=3.5cm]{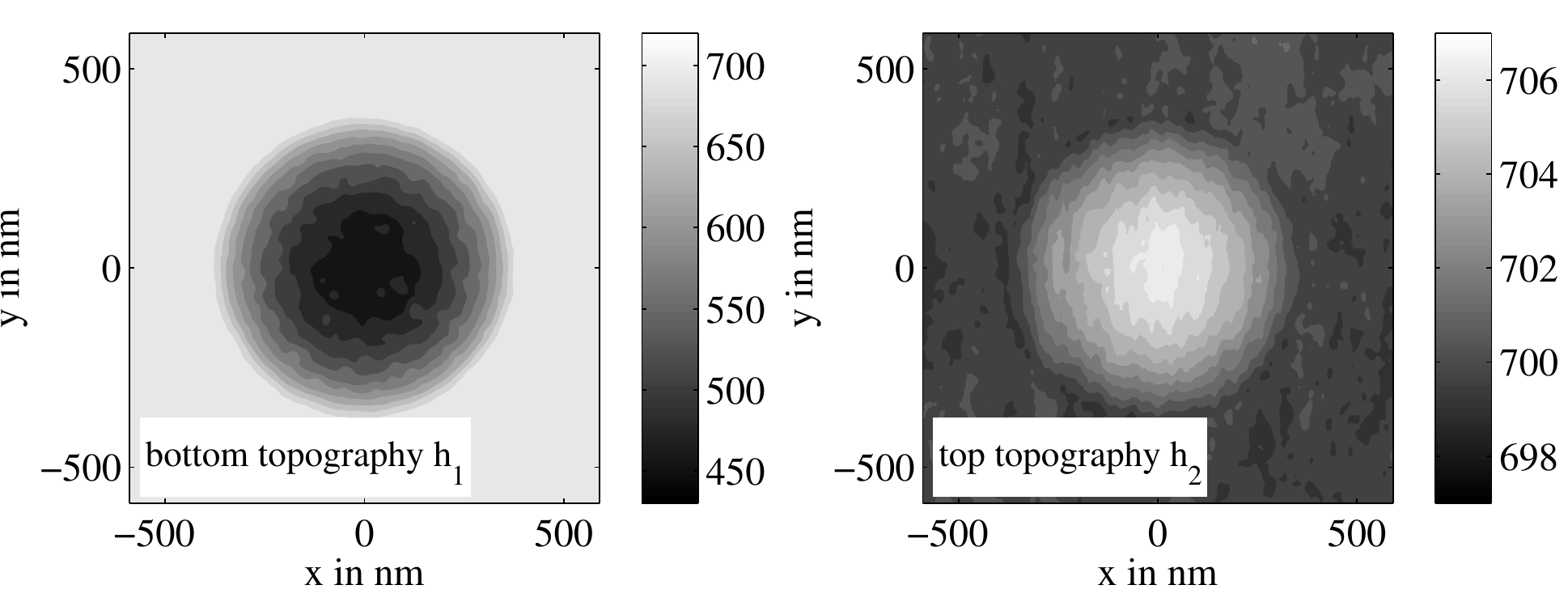}
\caption{Measured equilibrium drop profile as determined by AFM. The height scales in nanometers are shown to the right of each panel. (\emph{left}) bottom profile $h_1$ and (\emph{right}) top profile $h_2$.}
\label{fig:droplet1}
\end{figure}

\subsubsection*{Extraction of surface tension and contact angles}

Using the above validated protocol allows to measure three dimensional PS drop profiles on PMMA and to determine the contact angles and the ratios of the surface tensions from their equilibrium shapes. These values will serve as input parameters for the computed drop morphologies and could not be found in literature with the needed precision \cite{anastasiadis1988interfacial,wu1970surface}.

\begin{figure}
\centering
\includegraphics[width=8cm]{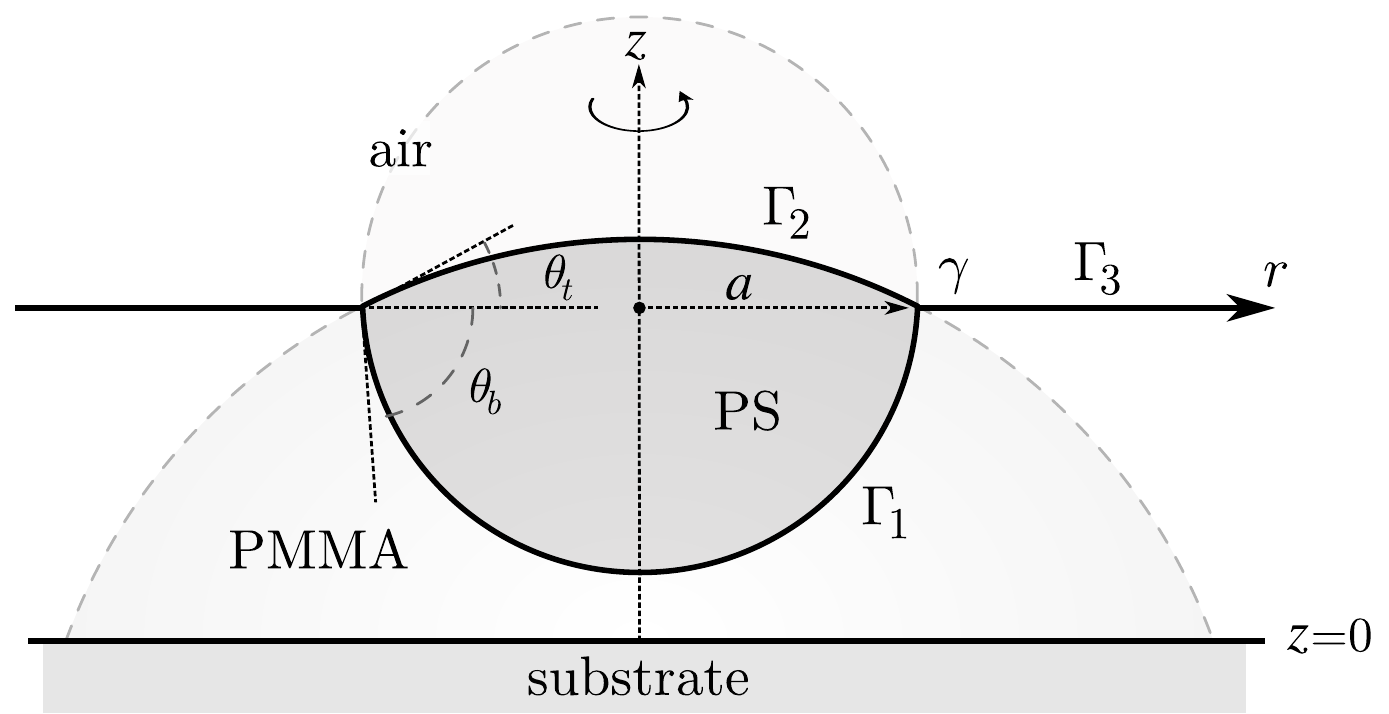}
\caption{Sketch of a axisymmetric liquid lens with $\Gamma_1$, $\Gamma_2$, $\Gamma_3$ being the PS/PMMA, PS/air, PMMA/air interface respectively and $\gamma$ the triple-junction. Each interface $\Gamma_i$ has the surface tension
$\sigma_i$. The radius of the lens is $a$ and the contact angles are $\theta_t,\theta_b$ (top, bottom). Furthermore for planar
axisymmetric droplets we have
 $\mathbf{n}_{\Gamma_1}=-\mathbf{e}_r\cos\theta_b-\mathbf{e}_z\sin\theta_b$,
 $\mathbf{n}_{\Gamma_2}=-\mathbf{e}_r\cos\theta_t+\mathbf{e}_z\sin\theta_t$,
 $\mathbf{n}_{\Gamma_3}=\mathbf{e}_r$.}
\label{fig:setup}
\end{figure}

The analysis of the equilibrium drop shapes is based on the following reasoning: Liquid equilibrium droplets with constant surface-tension $\sigma_\alpha$ deform an underlying liquid substrate to minimize the surface energy
\begin{align}
\label{eqn:energy}
E = \sum_{\alpha=\{1,2,3\}} \sigma_\alpha \int_{\Gamma_{\alpha}} d\Gamma
\end{align}
for fixed droplet volumes.  A classical calculation shows that minimizers of $E$ have contact angles determined by the well-known Neumann-triangle
\begin{equation}
\label{eqn:neumann_triangle1}
\sum_{\alpha=\{1,2,3\}} \sigma_\alpha \mathbf{n}_{\Gamma_\alpha}= 0,
\end{equation}
where $\sigma_\alpha>0$ corresponds to the surface tensions of the PS/PMMA, PS/air, 
PMMA/air interface 
for $\alpha=1,2,3$ and each interface $\Gamma_\alpha$ has constant
mean curvature $H_\alpha$.  The normalized vector $\mathbf{n}_\alpha$ is
tangential to the corresponding interface $\Gamma_\alpha$ and normal to the
contact line $\gamma$ as indicated in fig.~\ref{fig:setup}. Introducing the
spreading coefficient $\sigma=\sigma_3-\sigma_1-\sigma_2$ one can easily verify
that equilibrium droplets as in \eqref{eqn:neumann_triangle1} only exist if
$(-\sigma)>0$ and $\sigma_1,\sigma_2>\tfrac{1}{2}(-\sigma)$. A sketch of the 
equilibrium configuration is shown in fig.~\ref{fig:setup}.
From a measured equilibrium configuration we can thus extract the values for surface tension as follows: If for instance $\sigma_3$ is given, then one can determine the other two surface tensions from the contact angles by solving the linear equation
\begin{align}
\label{eqn:neumann_triangle2}
\begin{pmatrix}
\cos\theta_t & \cos\theta_b \\
-\sin\theta_t & \sin\theta_b
\end{pmatrix}
\begin{pmatrix}
\sigma_{2}\\
\sigma_{1}
\end{pmatrix}
=
\begin{pmatrix}
\sigma_{3}\\
0
\end{pmatrix}
\end{align}
where $\theta_t,\theta_b>0$ are measured. For contact angles $\theta_t,\theta_b\le 90\degree$ a liquid lens has the following equilibrium shape for $r \le a$
\begin{subequations}
\label{eqn:spherecap}
\begin{align}
h_2(x,y)&=h_\infty+\left(\sqrt{H_2^{-2}-r^2}-\sqrt{H_2^{-2}-a^2}\right)\\
h_1(x,y)&=h_\infty-\left(\sqrt{H_1^{-2}-r^2}-\sqrt{H_1^{-2}-a^2}\right)
\end{align}
\end{subequations}
and $h_1(x,y)=h_2(x,y)=h_\infty$ for $r>a$. We use the radial coordinate $r^2=(x-x_0)^2+(y-y_0)^2$ and call $a$ the radius of the droplet. We used the standard representation of interfaces $\Gamma_i$ by graphs of functions, \textit{i.e.}
\begin{subequations}
\begin{align}
\Gamma_1&=\{(x,y,z)\in\mathbb{R}^3:z=h_1(x,y),0<h_1<h_2\},\\
\Gamma_2&=\{(x,y,z)\in\mathbb{R}^3:z=h_2(x,y),0<h_1<h_2\},\\
\Gamma_3&=\{(x,y,z)\in\mathbb{R}^3:z=h_1(x,y),0<h_1=h_2\}.
\end{align}
\end{subequations}
Note that up to an additive constant basically $h_1(x,y)$ and $h_2(x,y)$ are the morphologies measured by the AFM, where $h_2$ is measured at $T=140\celsius$ and $h_1$ after the removal of the PS by an selective solvent.

A least-squares fit of \eqref{eqn:spherecap} to the measured AFM topography data returns the six parameters $h_\infty$, $H_1$, $H_2$, $a$, $x_0$ and $y_0$. Since both interfaces, \textit{i.e.} $h_1$ and $h_2$ are measured independently and the AFM can only measure height differences, $h_\infty$, $x_0$, $y_0$ have no absolute value, so one might define $x_0=y_0=0$ and $h_\infty$ as the values set by the preparation of the PMMA layer and as determined independently. Also $h_2$ is measured in its liquid state and $h_1$ after solidification and removal of the PS by a selective solvent. 
Using the values for $H_\alpha$ and $a$ the contact angles can be directly computed via
\begin{subequations}
\label{eqn:contactangle}
\begin{align}
\theta_b=\arctan \left({a/\sqrt{H_1^{-2}-a^2}}\right),\\
\theta_t=\arctan\left({a/\sqrt{H_2^{-2}-a^2}}\right).
\end{align}
\end{subequations}
Inserting this into \eqref{eqn:neumann_triangle2} gives the surface tensions.

\begin{figure}
\centering
\includegraphics[height=3.4cm]{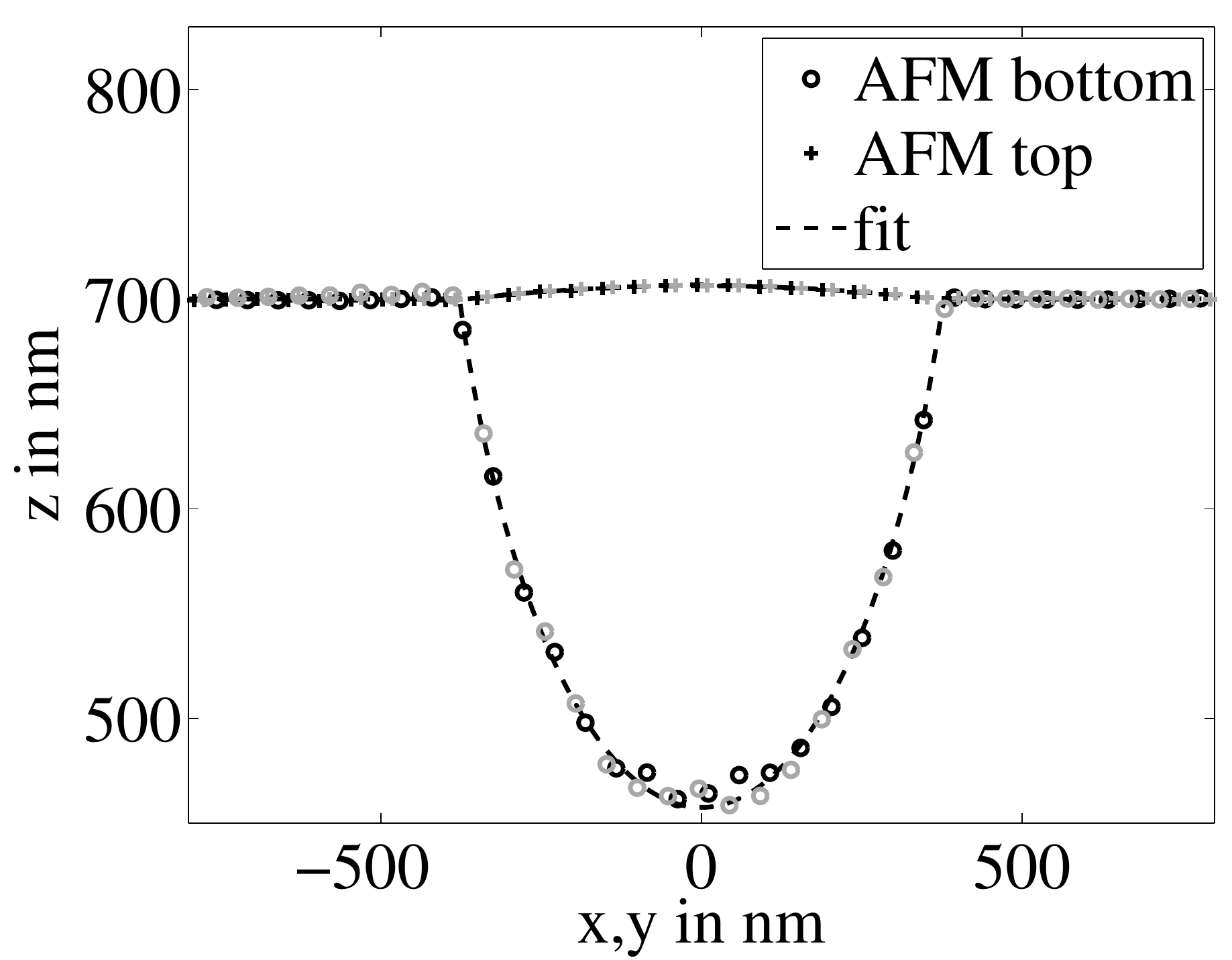}%
\includegraphics[height=3.4cm]{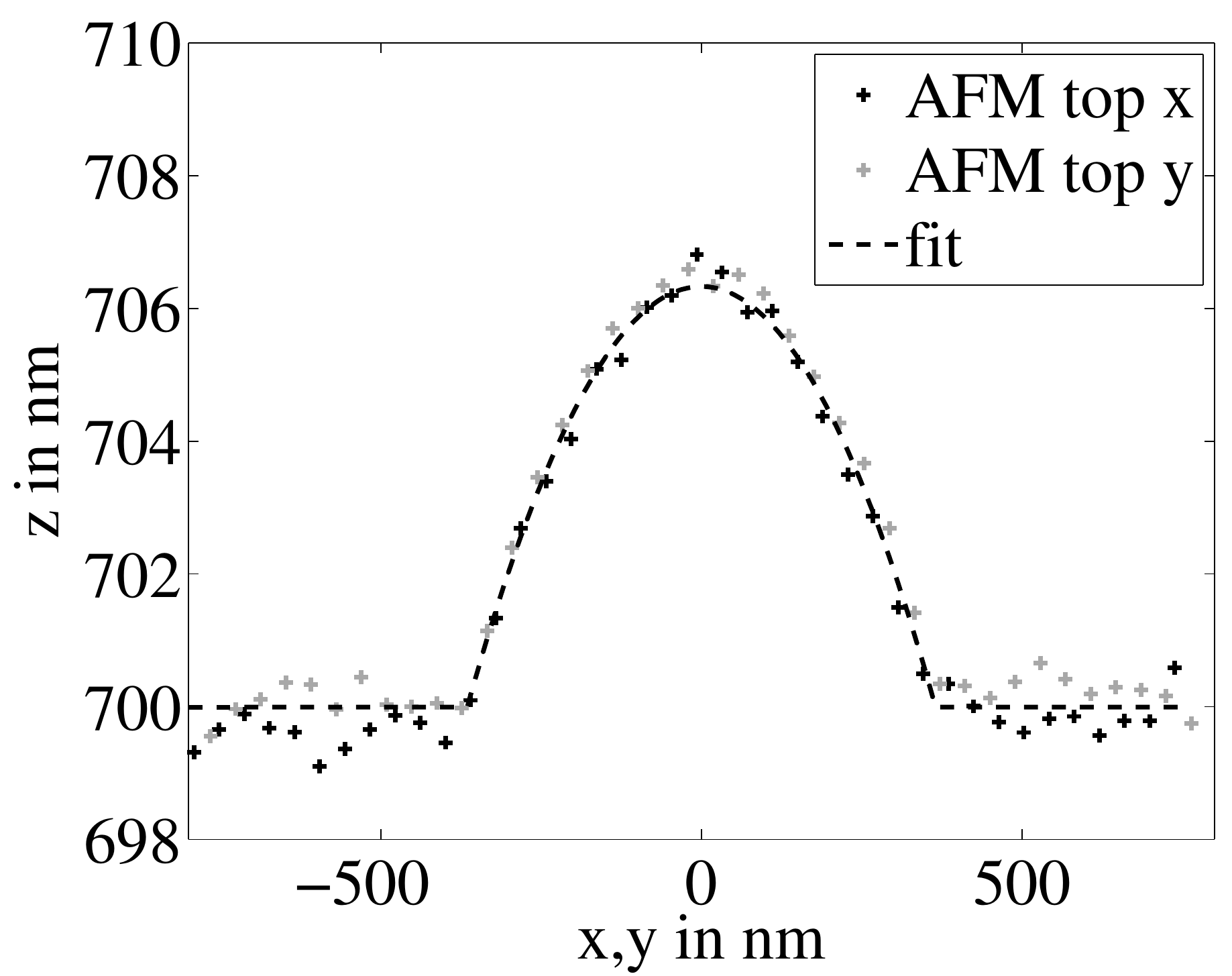}
\caption{(\emph{left}) Top and bottom AFM topography of an equilibrium droplet. Two cross sections cut perpendicular through the droplet in $x$-direction (dark symbols) and $y$-direction (light symbols) are shown together with the fit using eq.~\eqref{eqn:spherecap} (dashed line); (\emph{right}) close-up of the top AFM topography including fit function.}
\label{fig:droplet0}
\end{figure}

\begin{figure}
\centering
\includegraphics[width=.25\textwidth]{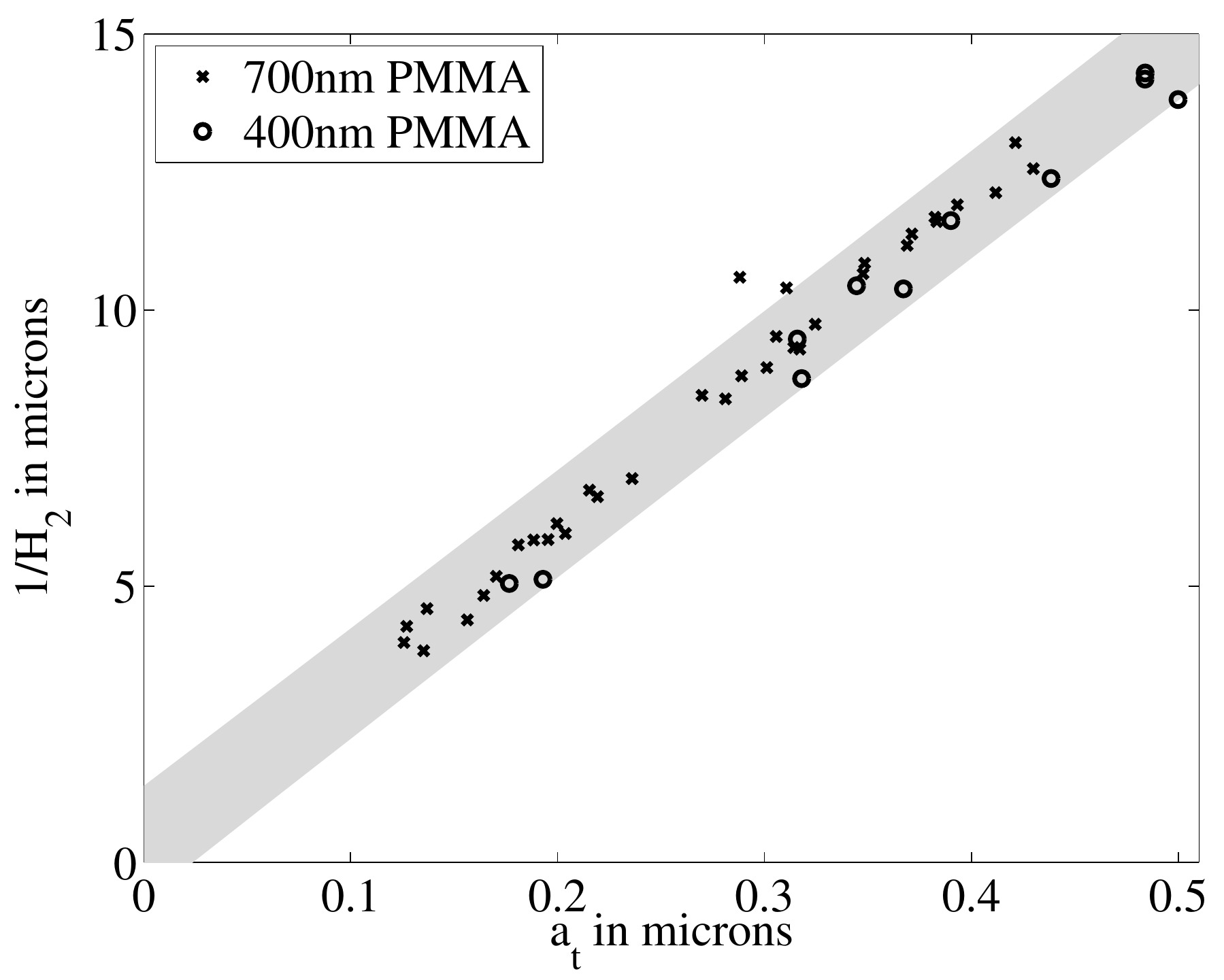}%
\includegraphics[width=.25\textwidth]{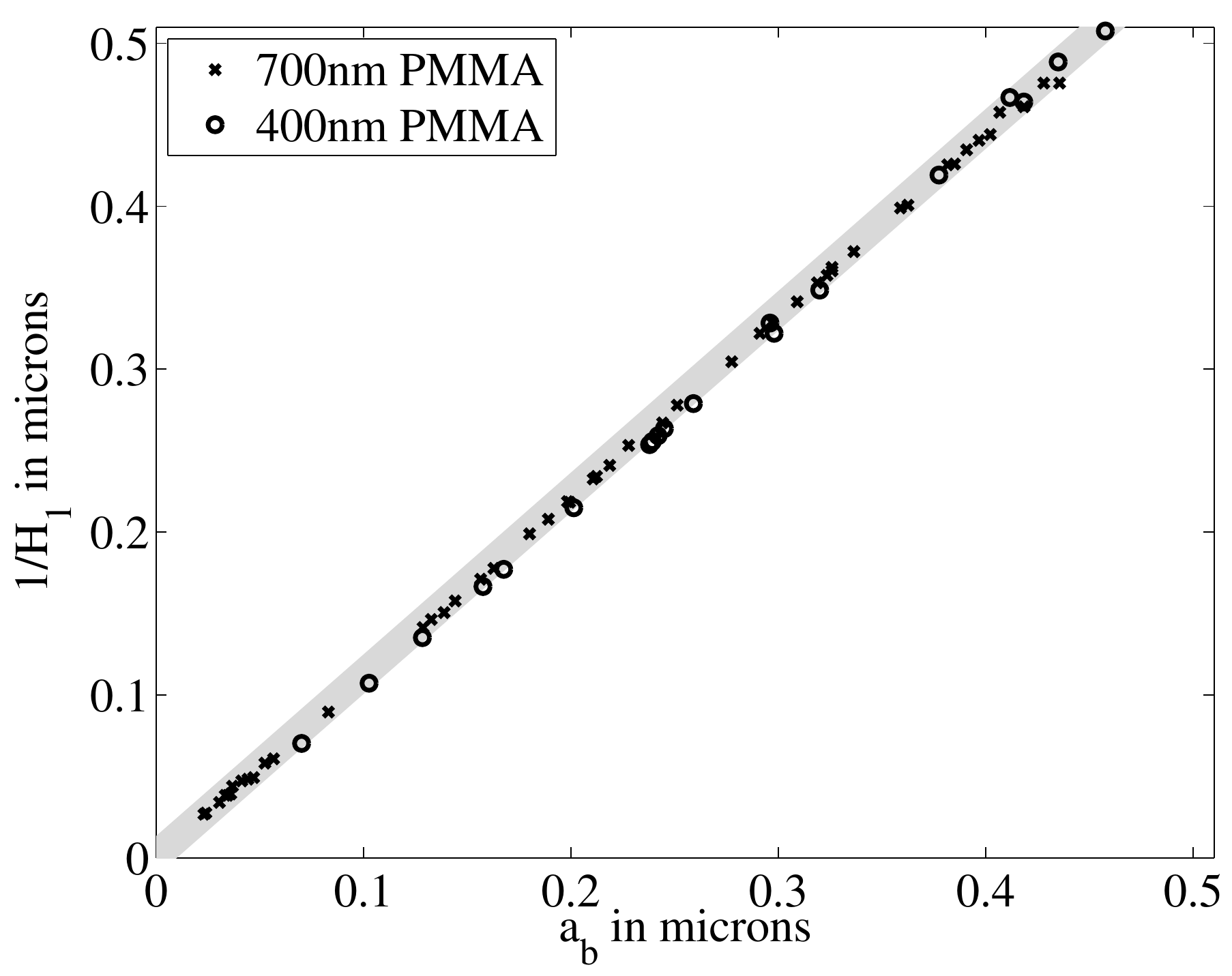}
\caption{(top) Curvature of the top spherical caps and (bottom) curvature of bottom spherical caps as a function of droplet radius $a$ measured from equilibrium droplets on $400\nm$ thick PMMA substrates (circles) and $700\nm$ thick PMMA substrates (crosses) with linear fit (shaded area $95\%$ confidence interval) gives $H_1^{-1}=(1.11\pm 0.02)a_b$ and $H_2^{-1}=(29\pm 1)a_t$ and corresponding
contact angles $\theta_b=(64\pm 2)\degree$ and $\theta_t=(1.98\pm 0.07)\degree$.}
\label{fig:curvcurv}
\end{figure}

Fitting spherical caps \eqref{eqn:spherecap} to the top and bottom profiles of several droplets, see fig.~\ref{fig:droplet0}, we obtain a relationship between $a_t,a_b$ and $H_1,H_2$, respectively which is shown in fig.~\ref{fig:curvcurv}. For constant contact angles \eqref{eqn:contactangle} suggests that the relationship between curvature and radius must be linear which agrees within the experimental accuracy with the experimental data.

Note that due to the magnitude of systematic  and statistical errors no evidence for a deviation from a linear relation in fig.~\ref{fig:curvcurv} could be observed.

From the linear relationship between radius and curvature shown in fig.~\ref{fig:curvcurv} and using \eqref{eqn:contactangle}
we obtain for the top angle $\theta_{\rm t}\,\approx\,(1.98\,\pm\,0.07)\degree$ and for the bottom angle $\theta_{\rm b}\,\approx\,(64\,\pm\,2)\degree$ where the error is composed from the statistical error in the fit and the systematic error in the determination of the droplet radius. Using \eqref{eqn:neumann_triangle2} we obtain the surface tensions of the PS/PMMA interfaces
\[\sigma_1 = (0.038 \pm 0.002)\sigma_3=(1.22\pm 0.07)\,\milli\newton/\meter,\]
and of the PS/air interface
\[\sigma_2  = (0.984 \pm 0.001)\sigma_3=(31.49\pm 0.03)\,\milli\newton/\meter,\]
based on the PMMA/air surface tension $\sigma_3=32\millinewton/\meter$ at $T=140\celsius$ taken from \cite{wu1970surface}. The corresponding spreading coefficient is $\sigma=(-0.022\pm 0.003)\sigma_3=(-0.7\pm 0.1)\millinewton/\meter$.
%
%

\subsection{Nonequilibrium droplet shapes}

Now we are using the parameters extracted from the equilibrium droplet shapes in order to plug them into a thin-film model and to investigate the evolution of droplets near equilibrium. Therefore we shortly explain how such a model is derived from a free boundary problem and how the corresponding partial differential equation is discretized using the method of finite differences. The question of fidelity of the lubrication approximation for the particular combination of surface tensions is beyond the scope of this paper and will be investigated in future. Since it is quite unrealistic that we can obtain sufficiently precise initial data from the experiment and use these for a numerical simulation, we first check the dependence of our results on a particular choice of initial data. Then we compare a set of simulations for a fixed initial PMMA layer thickness of $80\nm$ at a specific time with corresponding experiments at exactly the same physical time.

Let us consider the final evolution of PS droplets on PMMA as they approach their stationary state. The Stokes equation which describes the evolution of the free boundaries $\Gamma_1(t),\Gamma_2(t),\Gamma_3(t)$ by a velocity $\mathbf{u}$ or correspondingly the evolution of $h_1(t,x,y)$ and $h_2(t,x,y)$ has a dissipative structure with energy $E$ in \eqref{eqn:energy} and for Newtonian liquids the dissipation
\[
D=\sum_{i=1}^2\int_{\Omega_i} \frac{\mu_i}{2}(\nabla \mathbf{u} +
\nabla\mathbf{u}^\top)^2\,dx\,dy\,dz.
\]
After a standard lubrication approximation, where one assumes that the interfaces are shallow $\|\nabla h_1\|^2,\|\nabla h_2\|^2 =\mathcal{O}(\eps^2)$ for $\eps\ll 1$ a formal asymptotic calculation shows that $h_1(t,x,y),h_2(t,x,y)$ are solutions of a fourth-order, degenerate parabolic equation. A standard method to circumvent the degeneracy of that equation and to avoid the non-smoothness of the energies is to introduce a precursor using an energy
\begin{align}
\label{eqn:lubenergy}
E_{h_*}=\int \left[\tfrac{1}{2}\tfrac{\sigma_1}{\sigma_2}\|\nabla h_1\|^2 +
\tfrac{1}{2}\|\nabla h_2\|^2 + V_{h_*}\right]\,dx\,dy.
\end{align}
where $V_{h_*}=V_{h_*}(h_2-h_1)$ also has the interpretation of an intermolecular potential which encodes van-der-Waals interactions and the appropriate Neumann contact angle.

\noindent
This leads to the following system of degenerate parabolic equations
\begin{subequations}
\label{eqn:lubsystem}
\begin{align}
\partial_t h_i = \sum_{j=1}^2 \nabla\cdot \left( Q_{ij}\,\nabla \pi_j \right),
\qquad\qquad i=\{1,\,2\},
\end{align}
where divergence and gradient are component-wise and with mobility matrix
\begin{equation}
Q=\frac{1}{\mu}\begin{pmatrix} \tfrac{1}{3}h_1^3 &
\tfrac{1}{2}h_1^2(h_2-\tfrac{1}{3}h_1) \\ \tfrac{1}{2}h_1^2(h_2-\tfrac{1}{3}h_1)
&\tfrac{1}{3}(\mu-1)(h_2-h_1)^3+\tfrac{1}{3}h_2^3 \end{pmatrix}
\end{equation}
\end{subequations}
and $\mu=\mu_1/\mu_2$ is the ratio of PMMA to PS viscosity. The reduced pressures are defined $\pi_j=\delta E_{h_*}/\delta h_j$.  This expansion is only valid if $\eps=\sqrt{{(-\sigma)}/{\sigma_2}} \ll 1$. All quantities above are non-dimensional and scaled using
\begin{equation*}
[z]=H, \qquad [x]=[y]=\eps^{-1} H,\qquad [t]=\eps^{-4} H \frac{\mu_2}{\sigma_2}%
\end{equation*}
for some height scale $H=20\nm$ which we define in correspondence to the initial PS layer thickness. Using the values of the surface tension and known viscosities mentioned in the methods section we obtain $\eps=0.15$, $[x]=133\nm$, $[t]=2.1\sec$, $\mu=350$ and $\sigma_1/\sigma_2=0.039$. Note that in \cite{jachalski2012stationary} some of the authors studied stationary solutions of \eqref{eqn:lubsystem} and highlighted the connection between $E$ in \eqref{eqn:energy} and $E_{h_*}$ in \eqref{eqn:lubenergy} as an appropriate limiting
problem for shallow interfaces and as $h_*\to 0$.

For better readability we state the model in non-\-dimen\-sional form but all
results, \textit{e.g.} numerical solutions, are presented in dimensional form again. The
particular potential we use is $V_{h_*}(h)=2h_*^3/h^3-3h_*^2/h^2$ and typically
we use $h_*=1/64$. Note that with such a small $h_*$  it is only important that
$V_{h_*}$ assumes its unique minimum $V_{h_*}(h_*)=-1$ and $V_{h_*}\nearrow 0$
as $h \gg h_*$ and $0<h_*\ll 1$. Furthermore our particular $V_{h_*}$ satisfies
$V_{h_*}\to\infty$  as $h\to 0$. With larger $h_*$ instabilities and effects
related  to \textit{e.g.} spinodal dewetting will become more important. Note that for
the model reduction to be asymptotically correct we need $(-\sigma)=O(\eps^2)$
but $\sigma_1,\sigma_2,\sigma_3=O(1)$.  As we showed in the previous section we
have $0<(-\sigma)\ll \sigma_2$ but $(-\sigma)\sim \sigma_1$. Nevertheless, we
observed, by comparison to preliminary calculations based on the full Stokes
model that the lubrication approximation still yields quite accurate results. In
fact, for many similar thin-film problems, see \textit{e.g.} \cite{kriegsmann2003steady}
or \cite{MW05} lubrication approximation is still useful near or even beyond its
formal validity.

In order to solve the axisymmetric system of lubrication equations
\eqref{eqn:lubsystem} we use a finite difference scheme with spatial and
temporal adaptivity. The spatial adaptivity is of second order and heuristically
adds points near the would-be triple junction, where second derivatives become
large as $h_*\to 0$. The time discretization uses a fully implicit Euler method
and the temporal adaptivity is achieved with standard step-size bisection to
control the error. With the exception of interpolation during a mesh adaptation
the scheme conserves the mass $m_i=2\pi\int_0^R h_i(t,r) r\,dr$. Boundary
conditions at $r=R$ and $r=0$ are Neumann conditions for $h_i$ and $\pi_i$.
As initial data for our comparison we use
\begin{subequations}
\label{eqn:idata}
\begin{align}
h_1(t=0,r)&=4H,\\
h_2(t=0,r)&=\begin{cases}4H + Hh_* + h & r<r_0 \\ 4H + Hh_* & r\ge r_0\end{cases}
\end{align}
\end{subequations}
where the initial disc radii $r_0$ and heights $h$ are choosen to match the volumes
and layer thickness prepared in the experiments.

\begin{figure}
\centering
\includegraphics[width=0.24\textwidth]{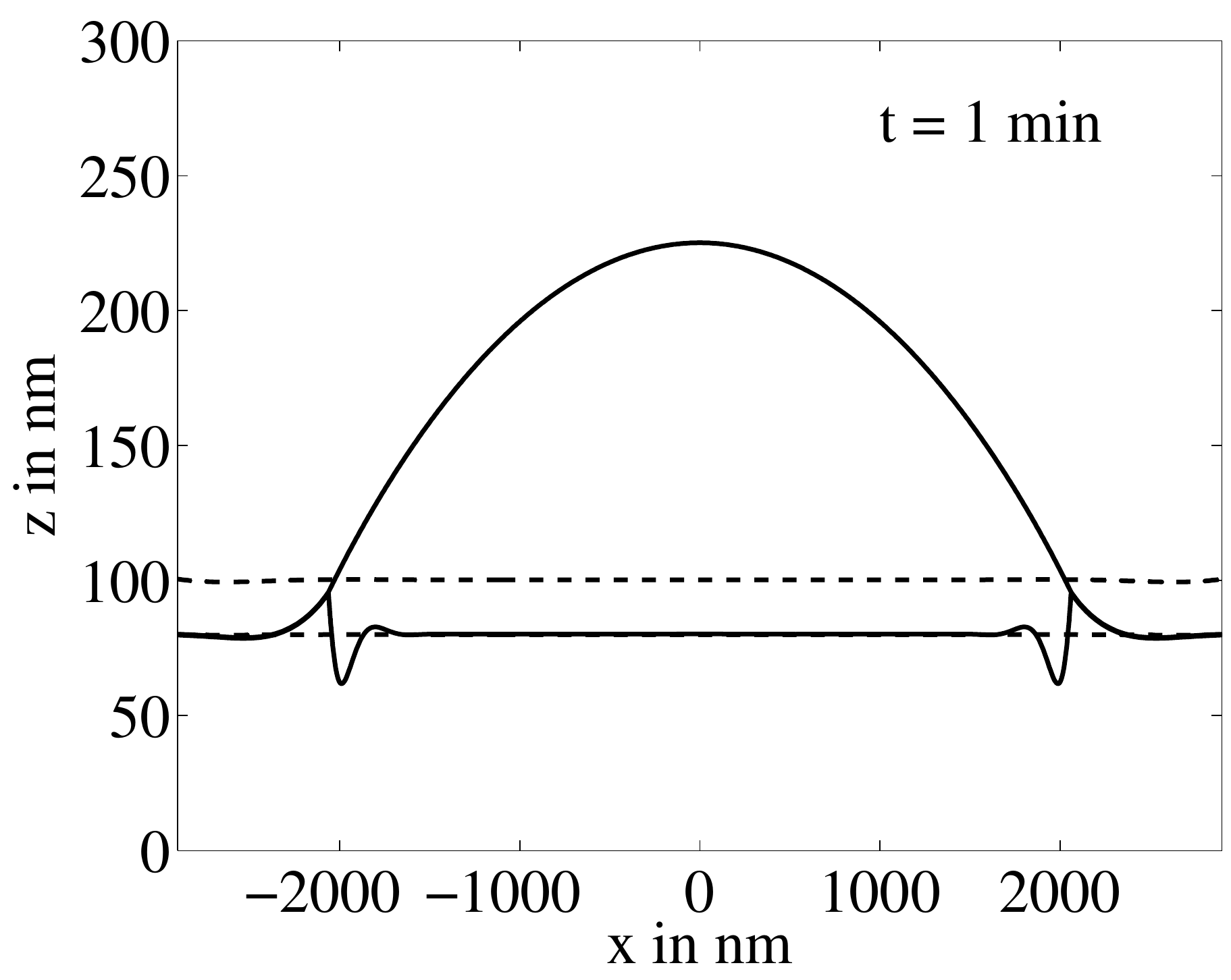}%
\includegraphics[width=0.24\textwidth]{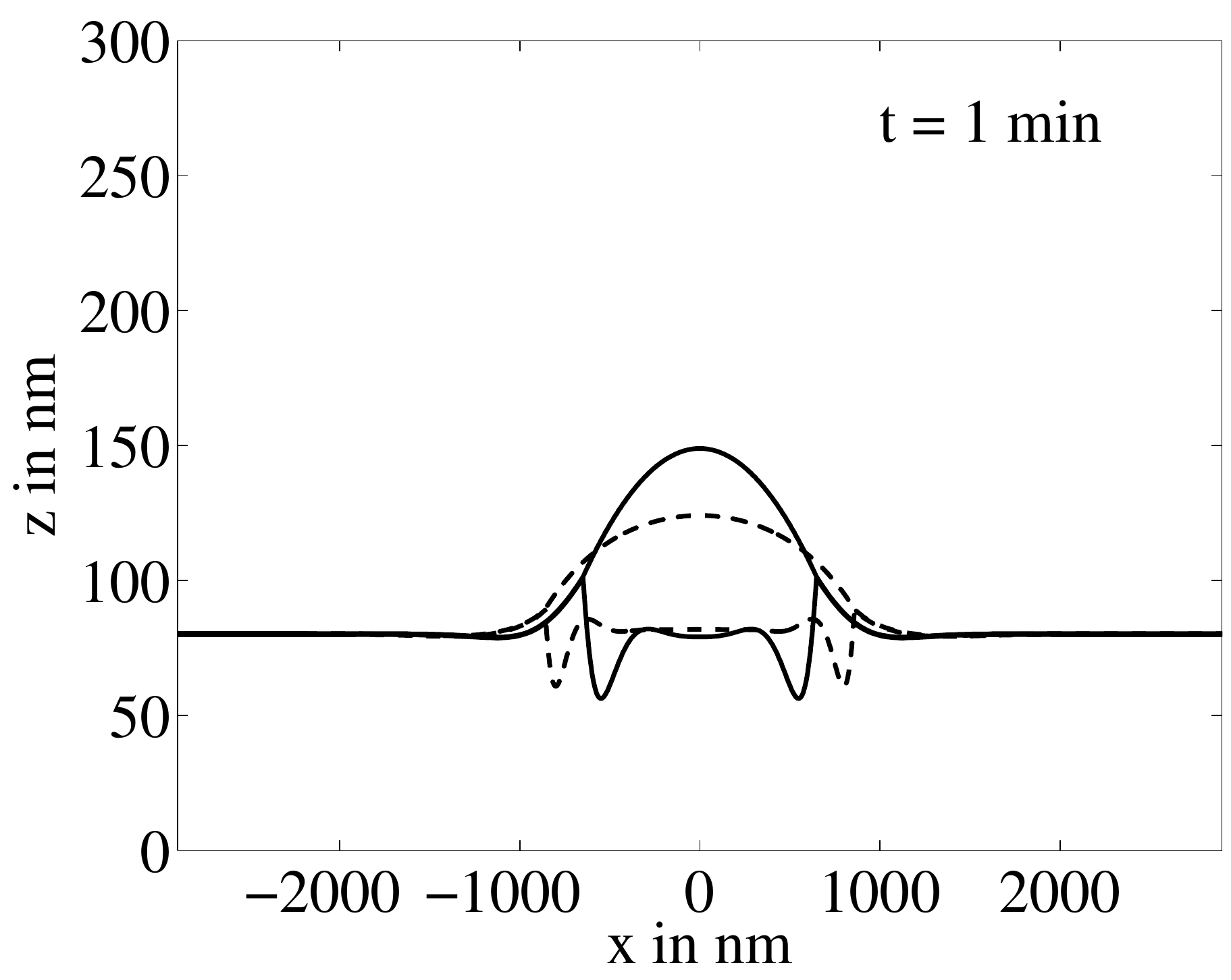}%

\includegraphics[width=0.24\textwidth]{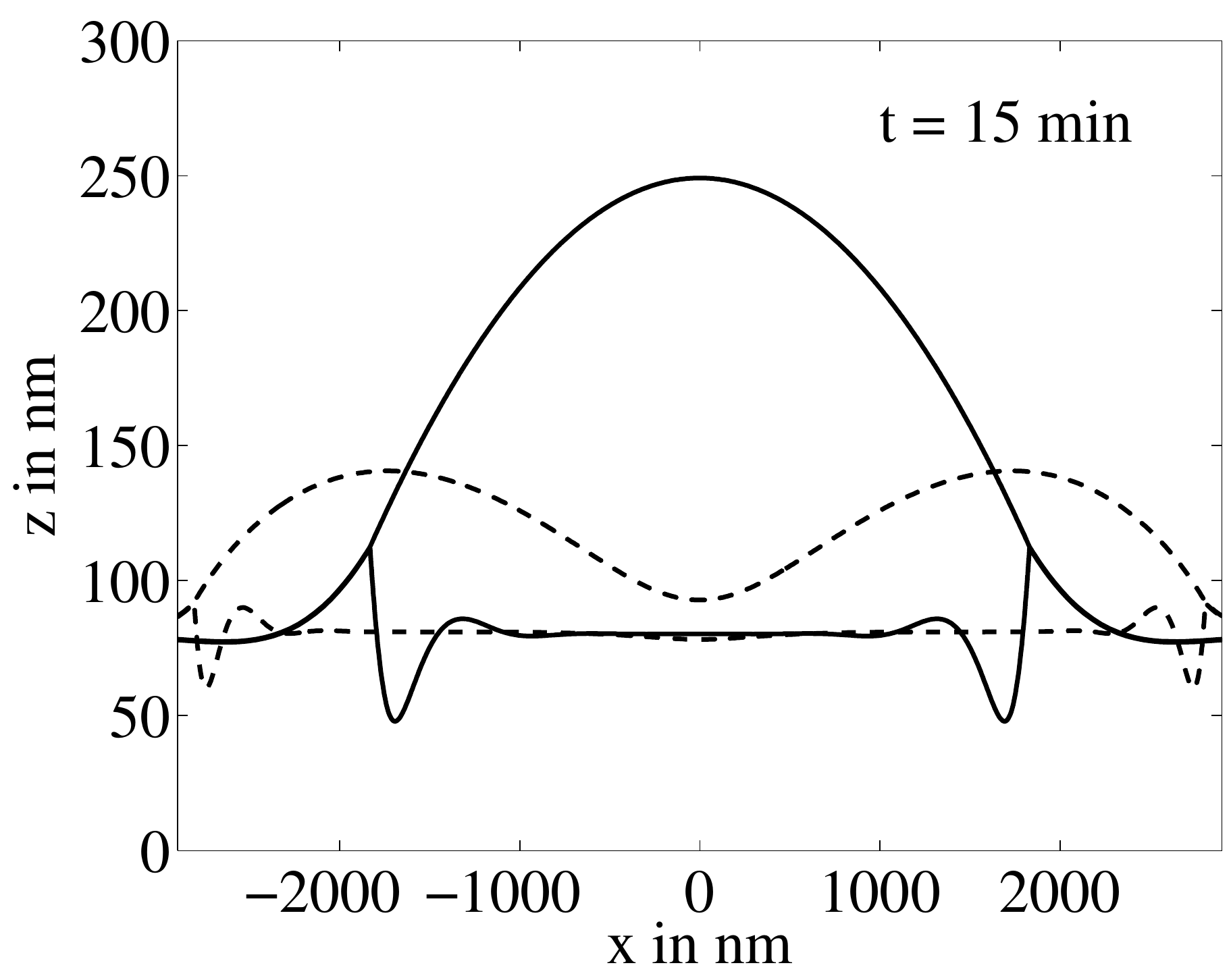}%
\includegraphics[width=0.24\textwidth]{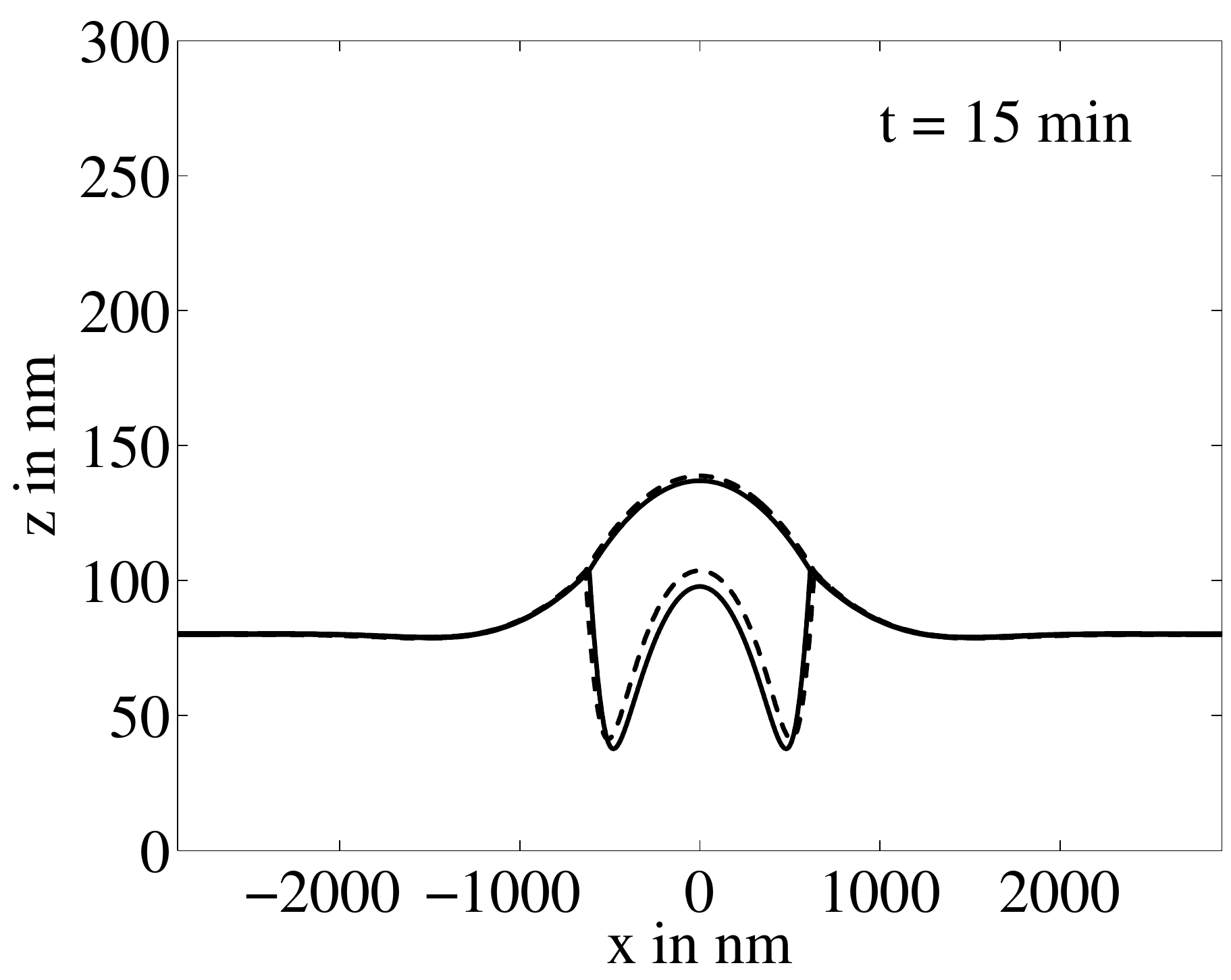}%

\includegraphics[width=0.24\textwidth]{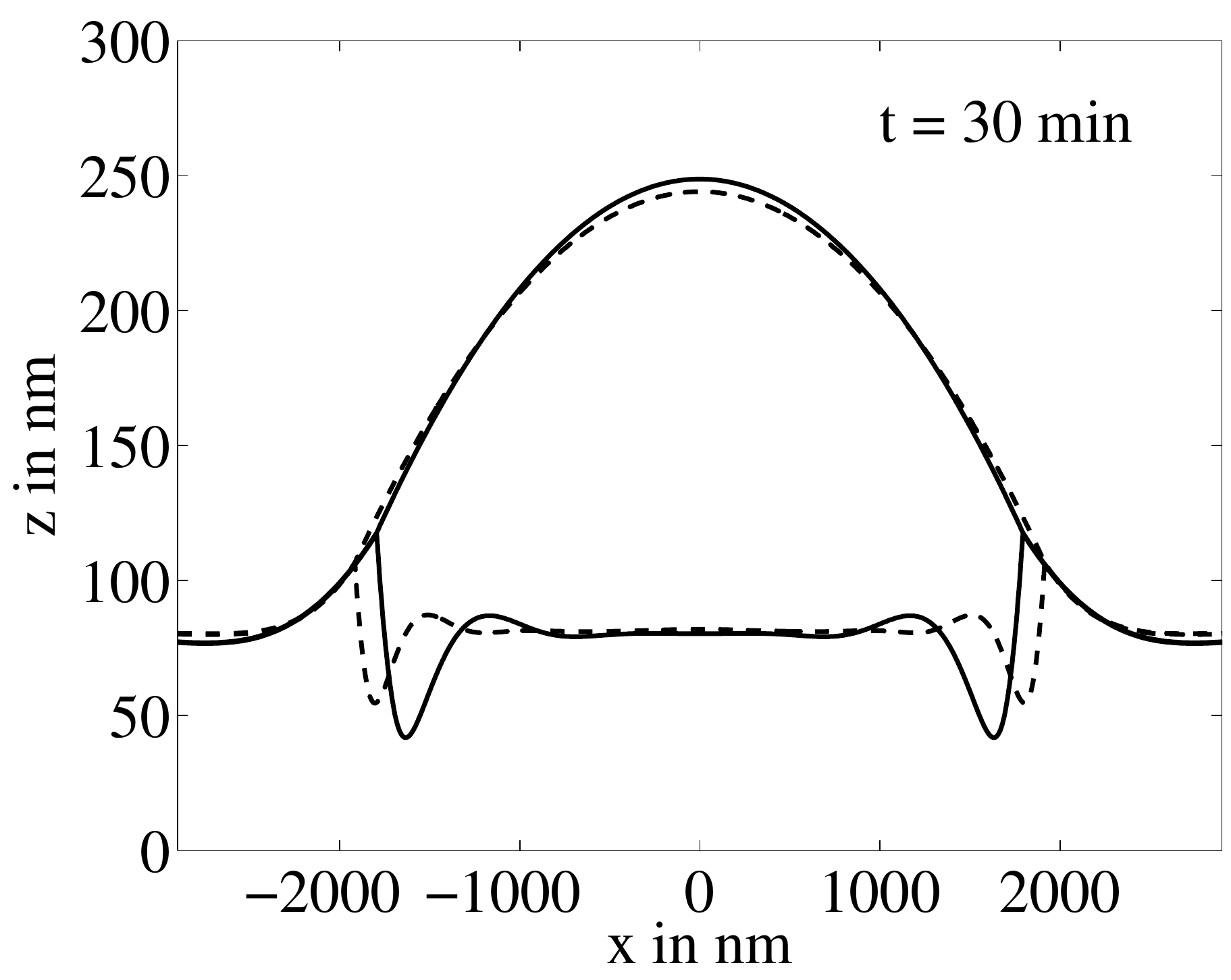}%
\includegraphics[width=0.24\textwidth]{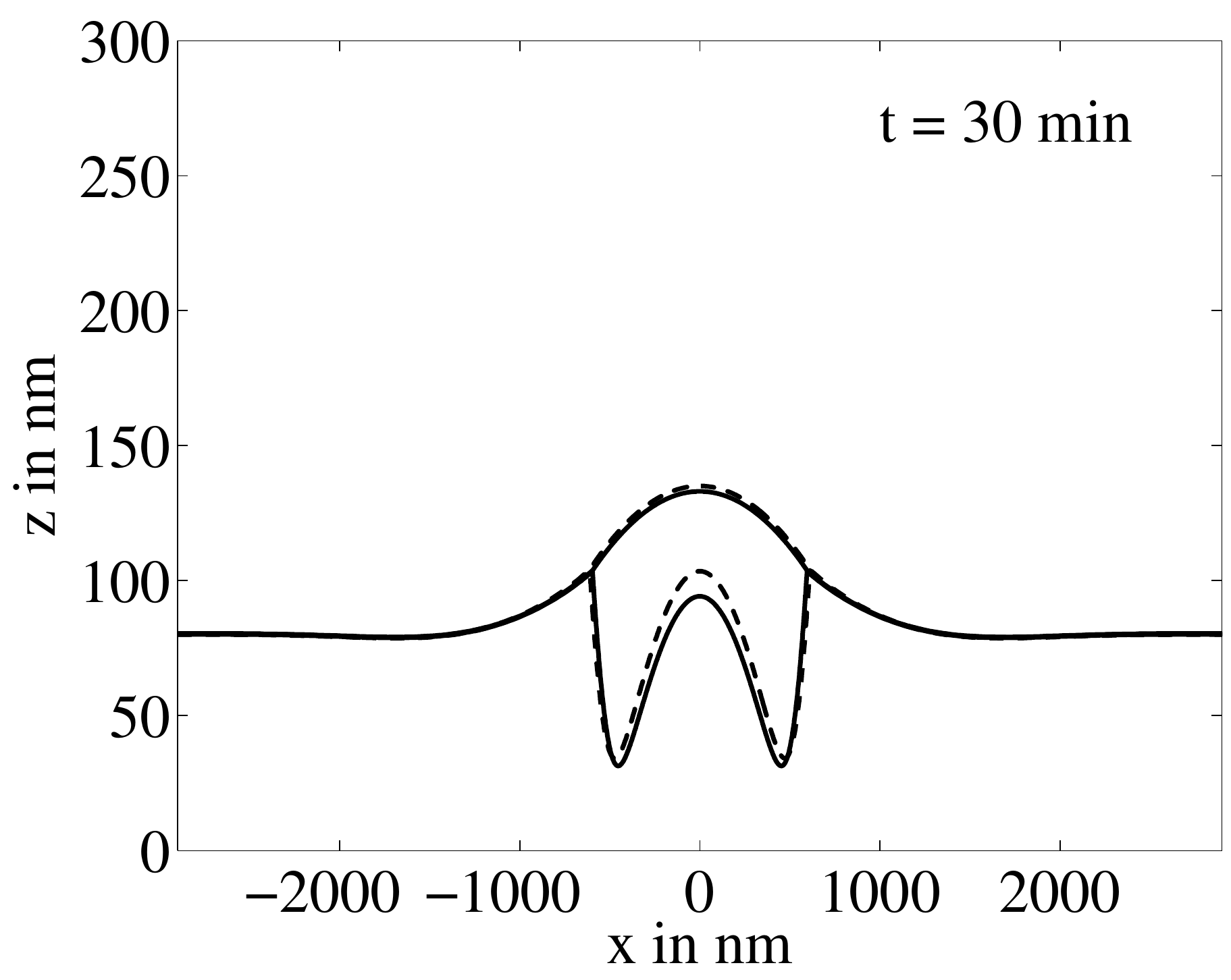}%

\includegraphics[width=0.24\textwidth]{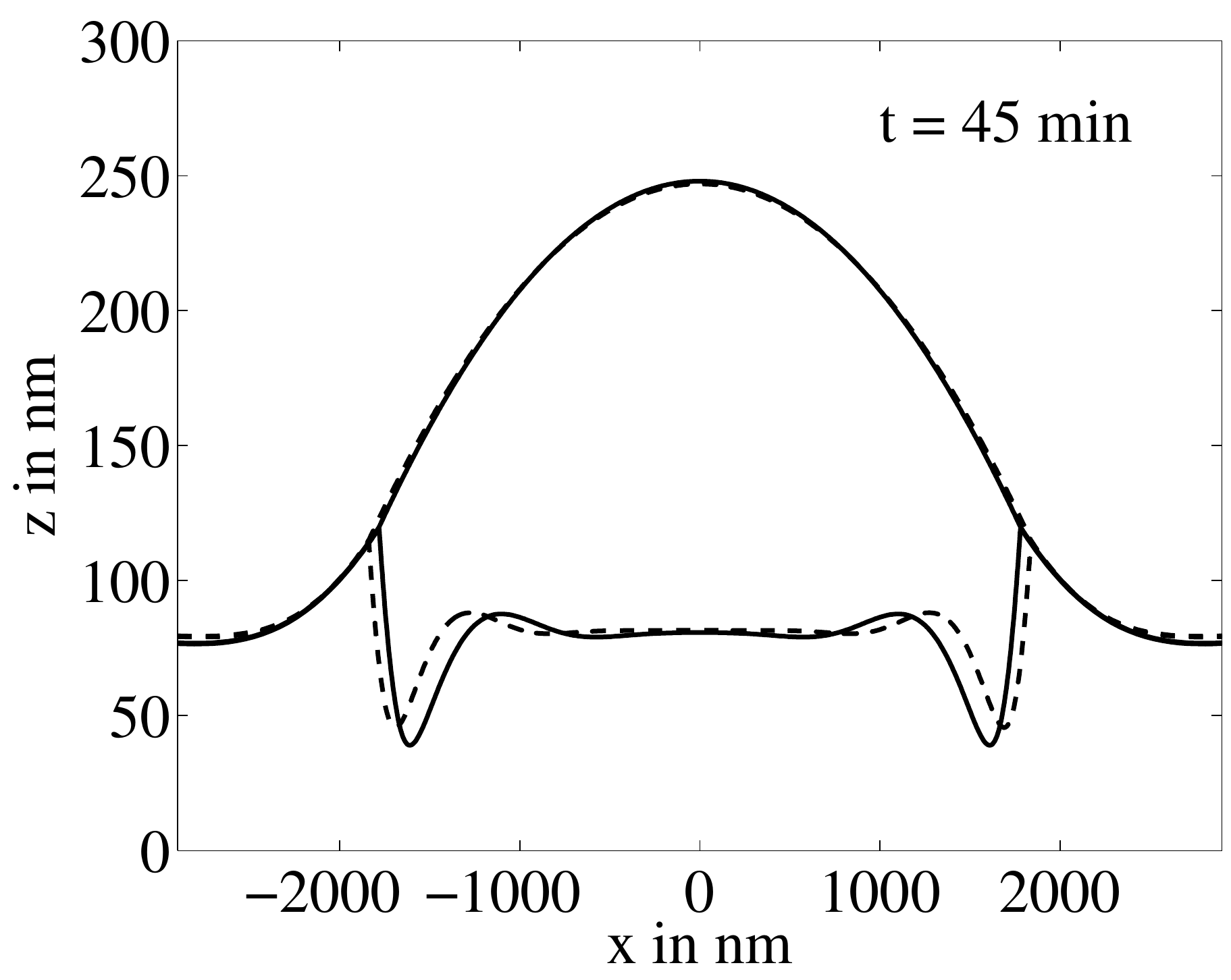}%
\includegraphics[width=0.24\textwidth]{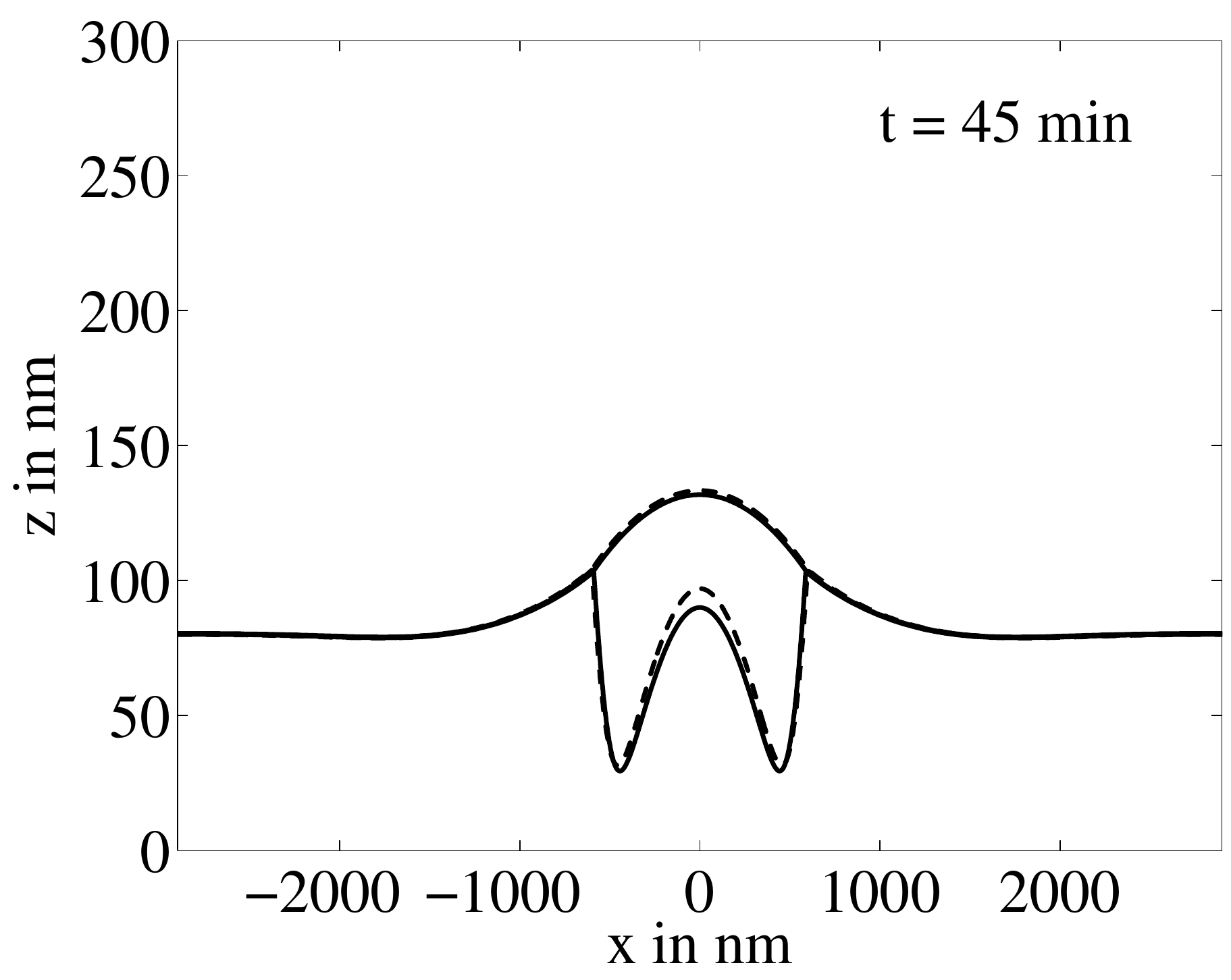}%

\caption{Numerical simulation of droplet formation and equilibration for two PS volumes and different initial configurations at different times. (left) $r_0=32L$, $h=H$ and $r_0=16L$, $h=4H$ and
(right) $r_0=8L$, $h=H$ and $r_0=4L$, $h=4H$ ($H=20\nanometer, L=133\nanometer$)}
\label{fig:evolidata}
\end{figure}

Now we consider numerical solutions of the thin film equation and their evolution towards equilibrium. In particular we fist show numerical solutions with two different volumes $2\pi \int (h_2-h_1) r\,dr=\{1.14,0.07\}\cubicmicrometer$. For each volume we use two different initial configurations as in \eqref{eqn:idata}. The strong volume dependence of the approach to equilibrium is demonstrated in  fig.~\ref{fig:evolidata}, where a synchronization of the different solutions can be observed as time progresses to infinity.

\begin{figure}
\centering
\includegraphics[height=0.17\textwidth]{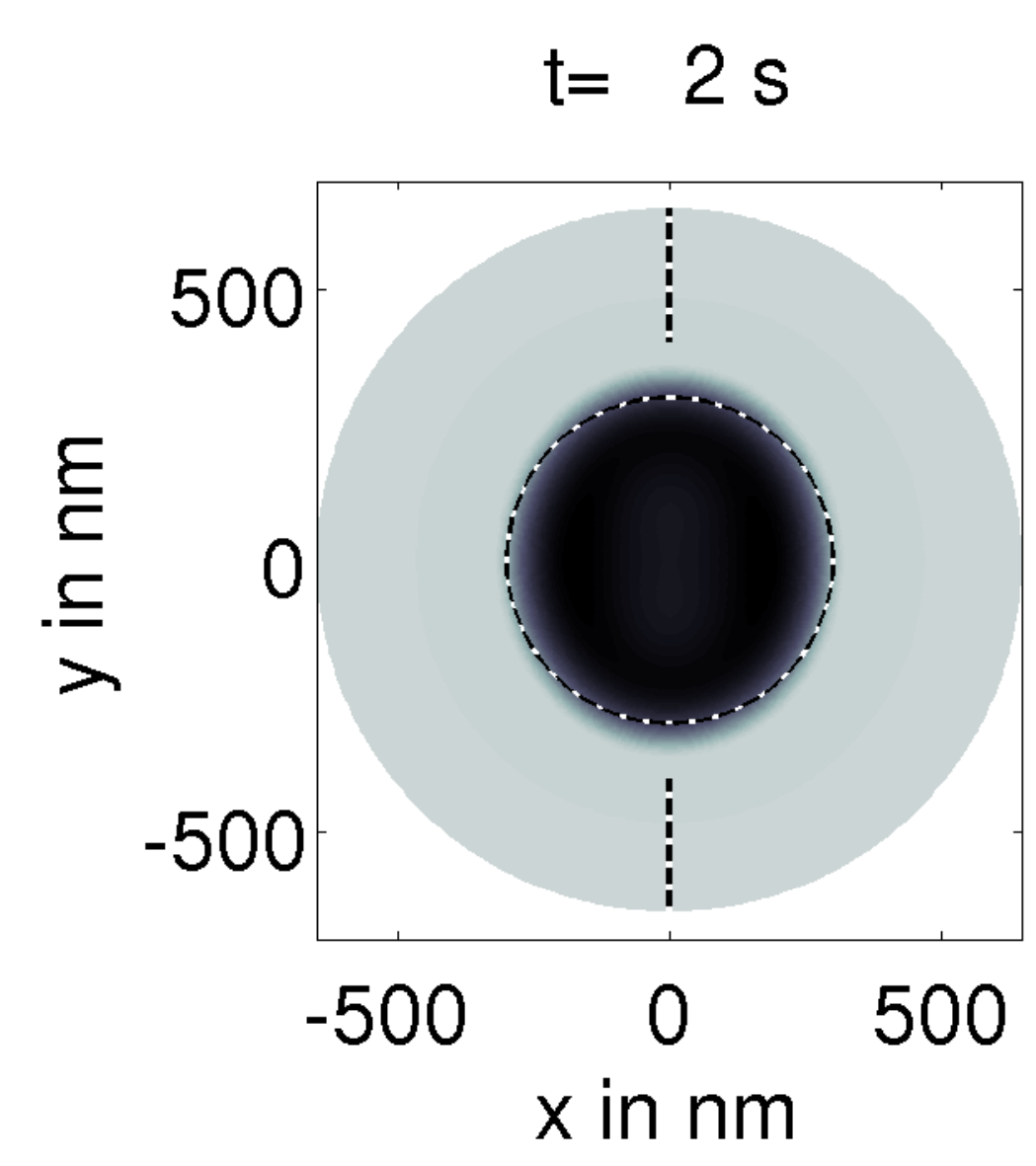}%
\includegraphics[height=0.17\textwidth]{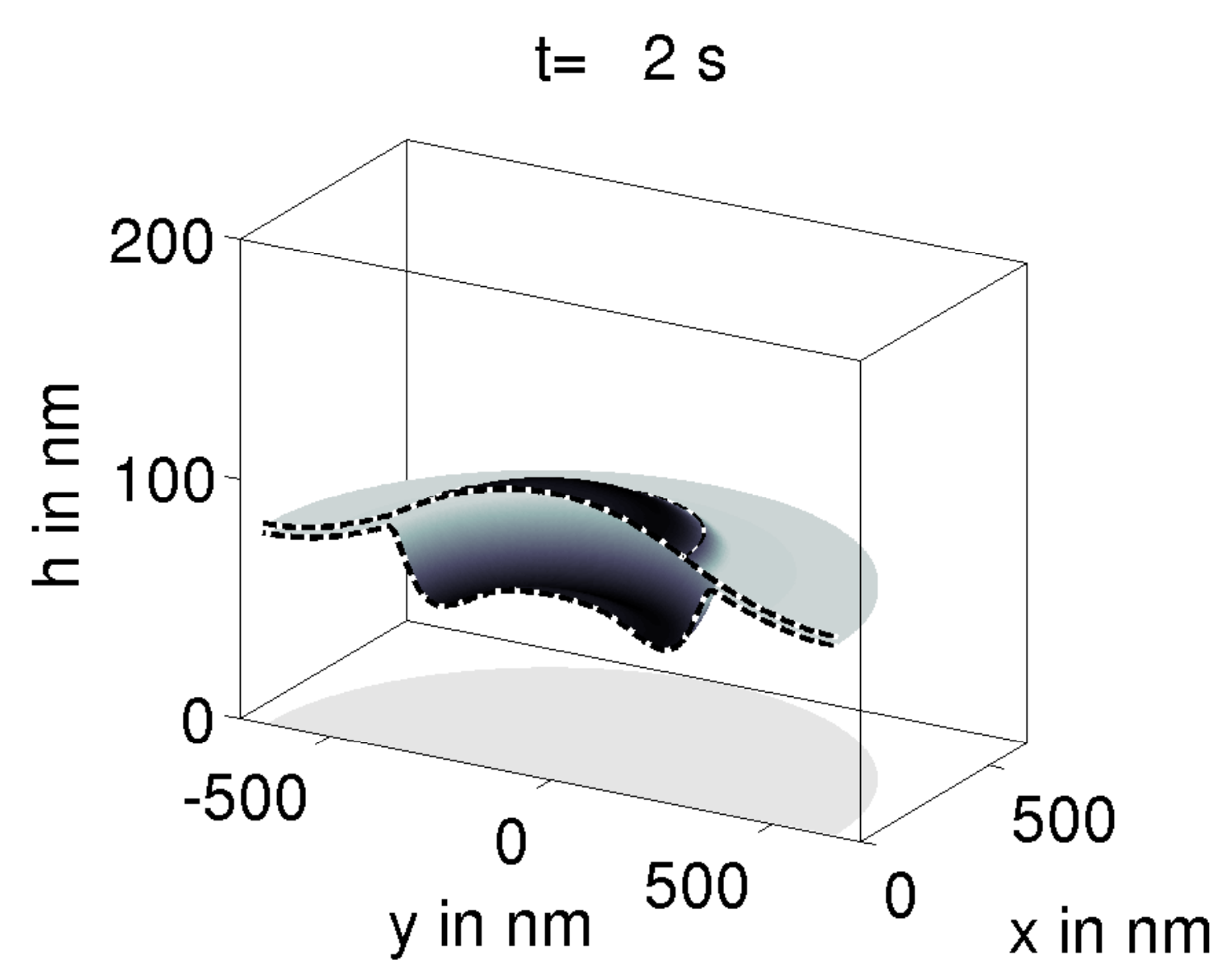}%

\includegraphics[height=0.17\textwidth]{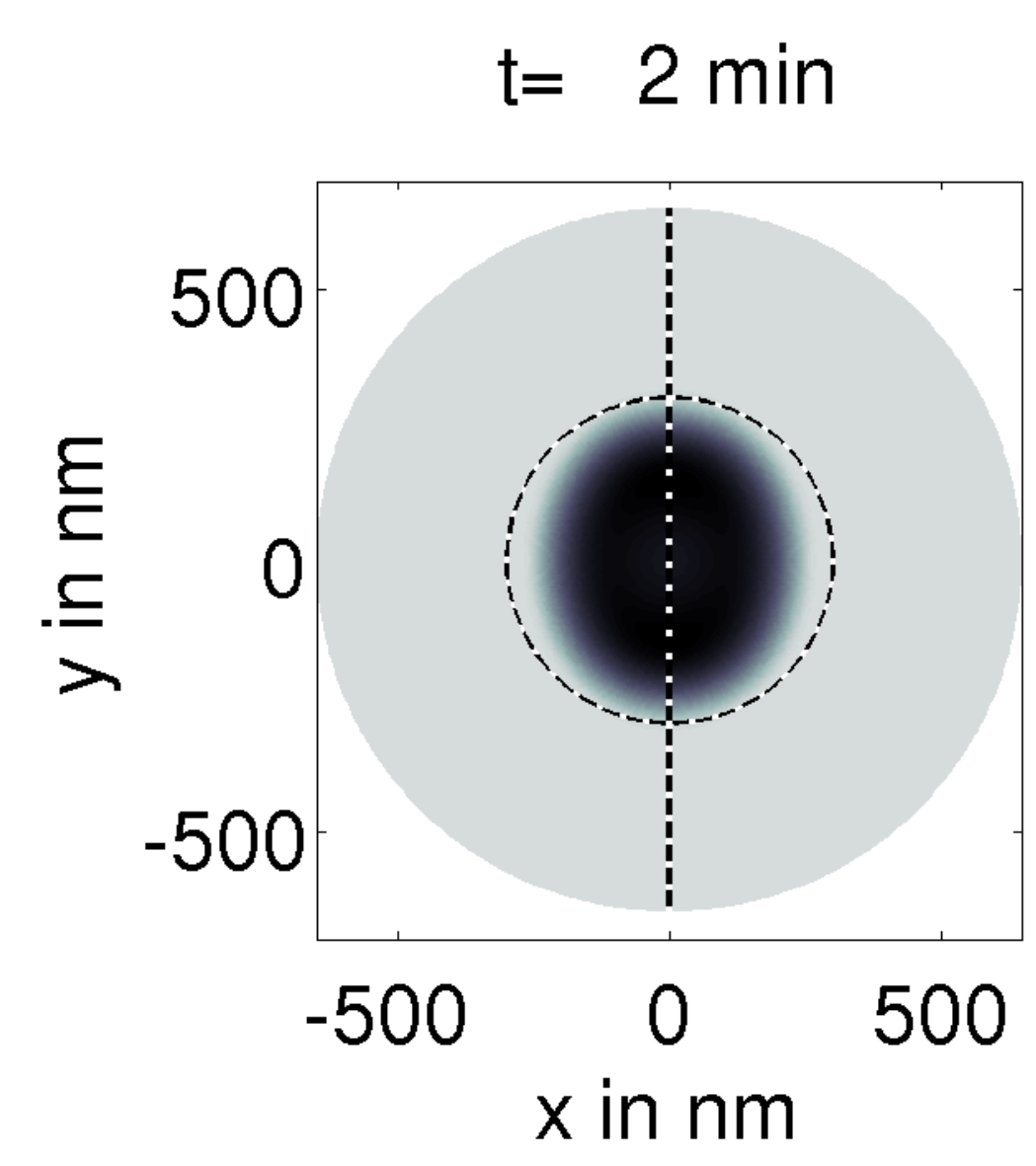}%
\includegraphics[height=0.17\textwidth]{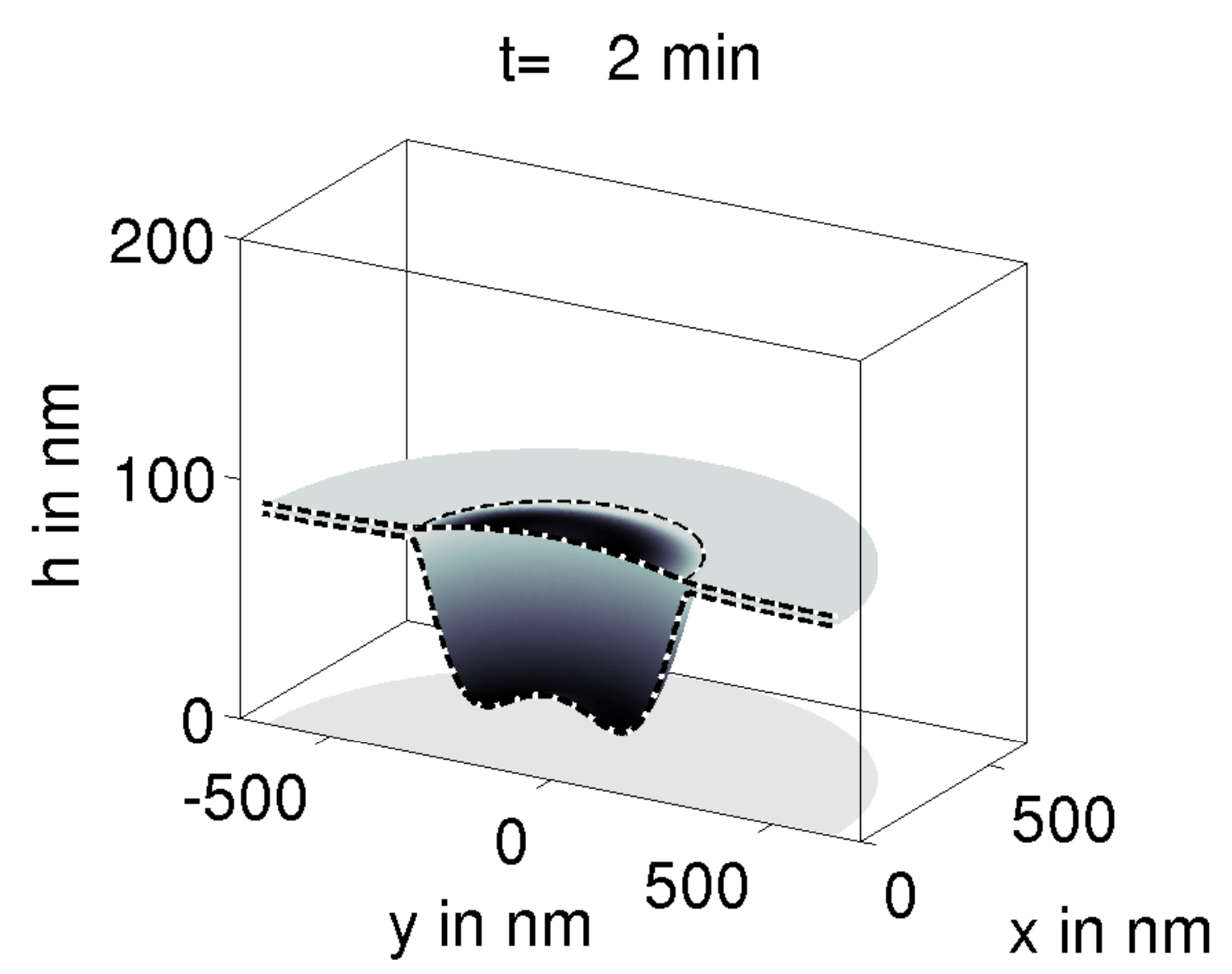}%

\includegraphics[height=0.17\textwidth]{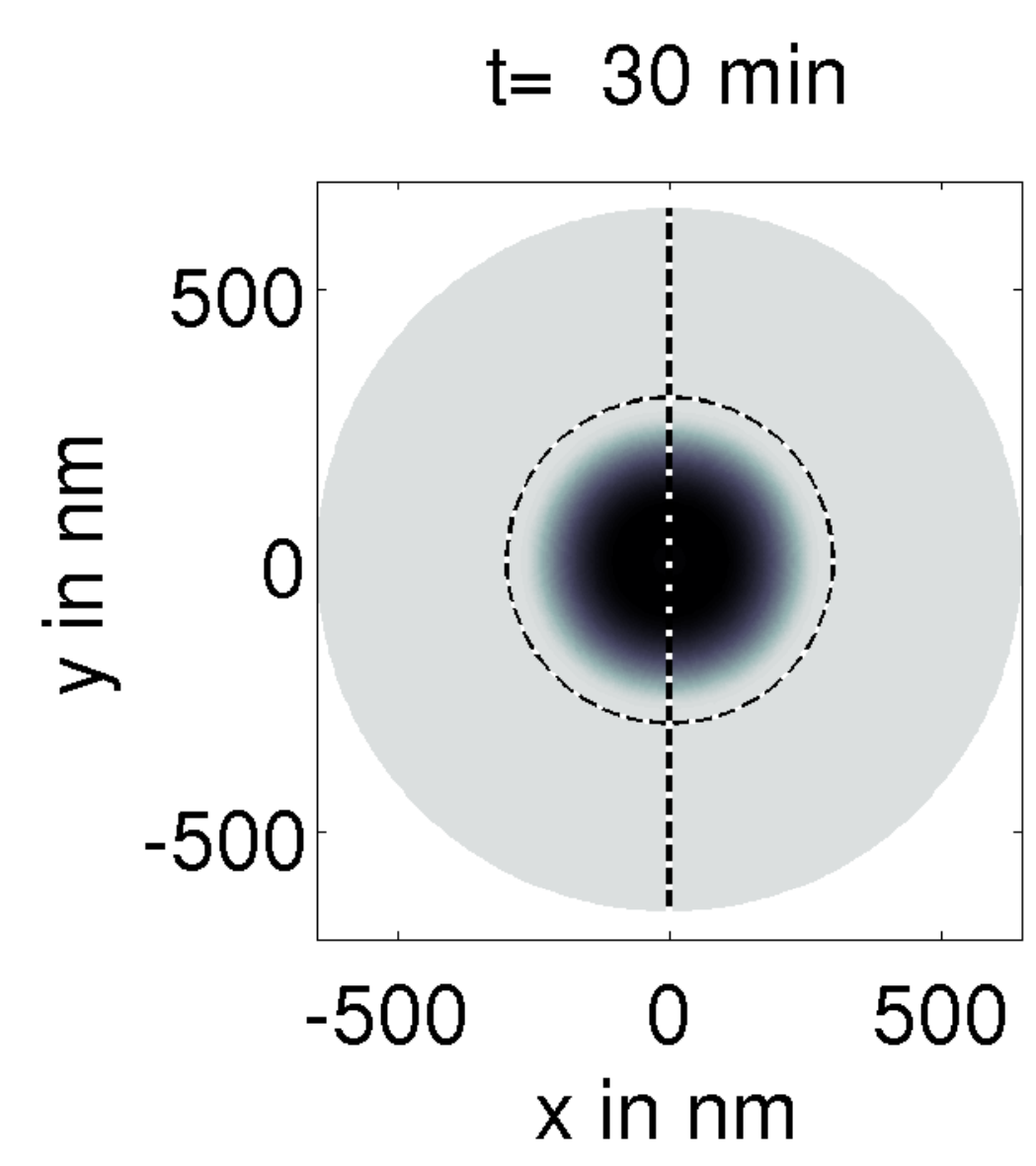}%
\includegraphics[height=0.17\textwidth]{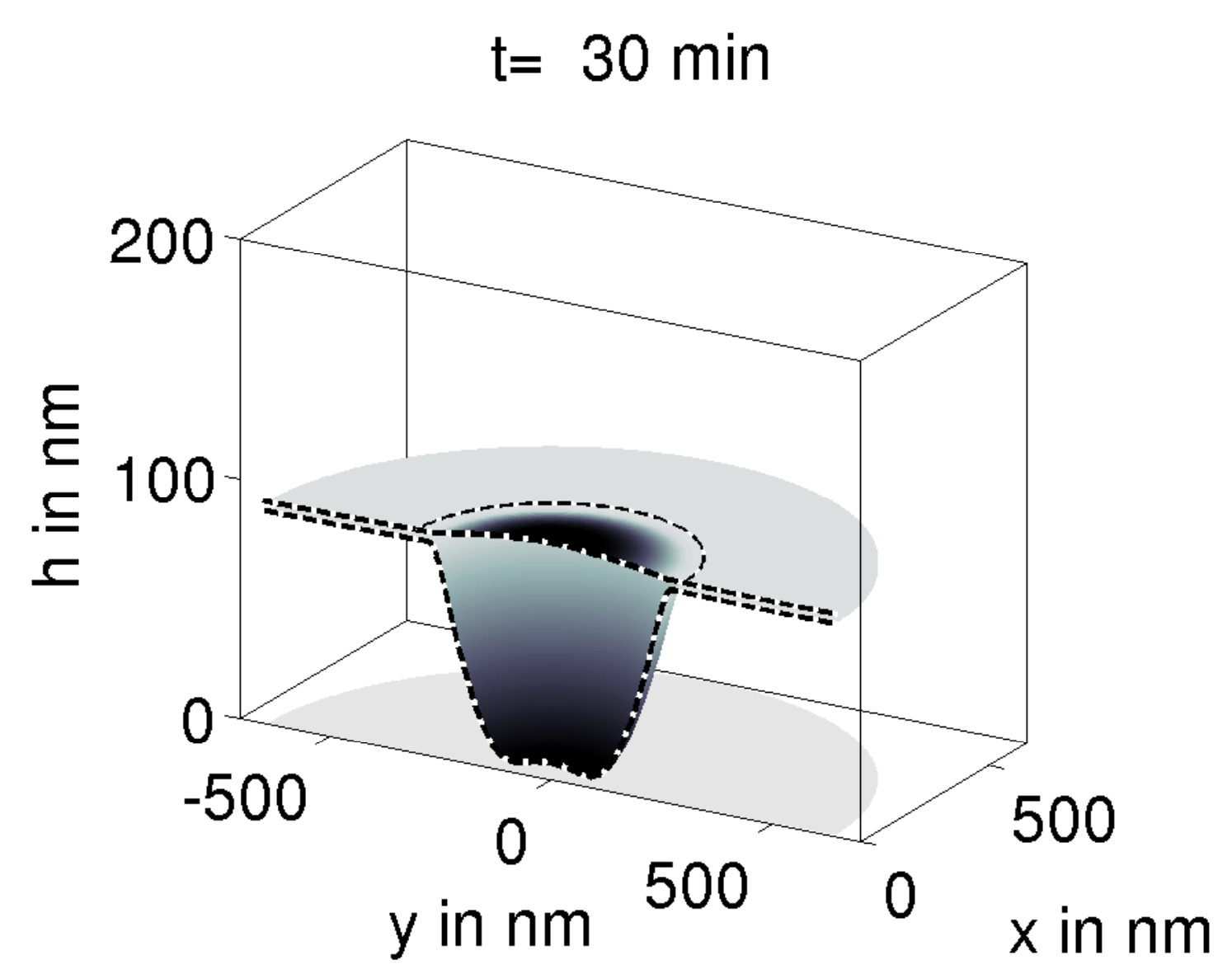}%
\caption{Numerical solutions of the $2d$ thin-film equation with nonsymmetric initial data and $h_*=1/5$ and other parameters as before at $t=2\sec,2\minute,30\minute$ (\emph{left}) top view (\emph{right}) 3d view showing a quick evolution into a nearly  axisymmtric state followed by an  evolution where $h_1\to 0$.}
\label{fig:latestage_nonsym}
\end{figure}

For the biggest droplet in the first row of fig.~\ref{fig:evolidata} this happens only after $45\min$,
while for the smallest droplets in the lower row different initial data are synchronized after less
than $1\min$. In our numerical simulations we have also started with asymmetric initial shapes and observe a quick approach to symmetrical shapes before the intermediate stages towards equilibrium have begun, see for illustration fig.~\ref{fig:latestage_nonsym}

\subsubsection*{Experiments vs thin-film model}

Using the precise knowledge of the surface tensions of our system and the weak dependence on initial data we are now able to follow the complete dewetting process to understand the stages of equilibration of a droplet by comparing these stages in experiment and simulation. For a given measurement at a certain time there is an upper bound in droplet size above which droplets are not
axisymmetric with their shape strongly dependent on initial data. In practice this implies that for the used liquid/liquid system of PS$(10k)$ and PMMA$(10k)$ at $T=140\celsius$ and realistic experimental time scales of a few days we are limited to droplets
of a few micrometers in radius. For experimental reason, as explained in the methods section, it is not possible to follow the liquid-liquid interface of a single droplet on its way to equilibrium. Thus different droplet volumes were imaged at the same dewetting time and compared to the related simulations.

\begin{figure}
\centering
\includegraphics[width=0.24\textwidth]{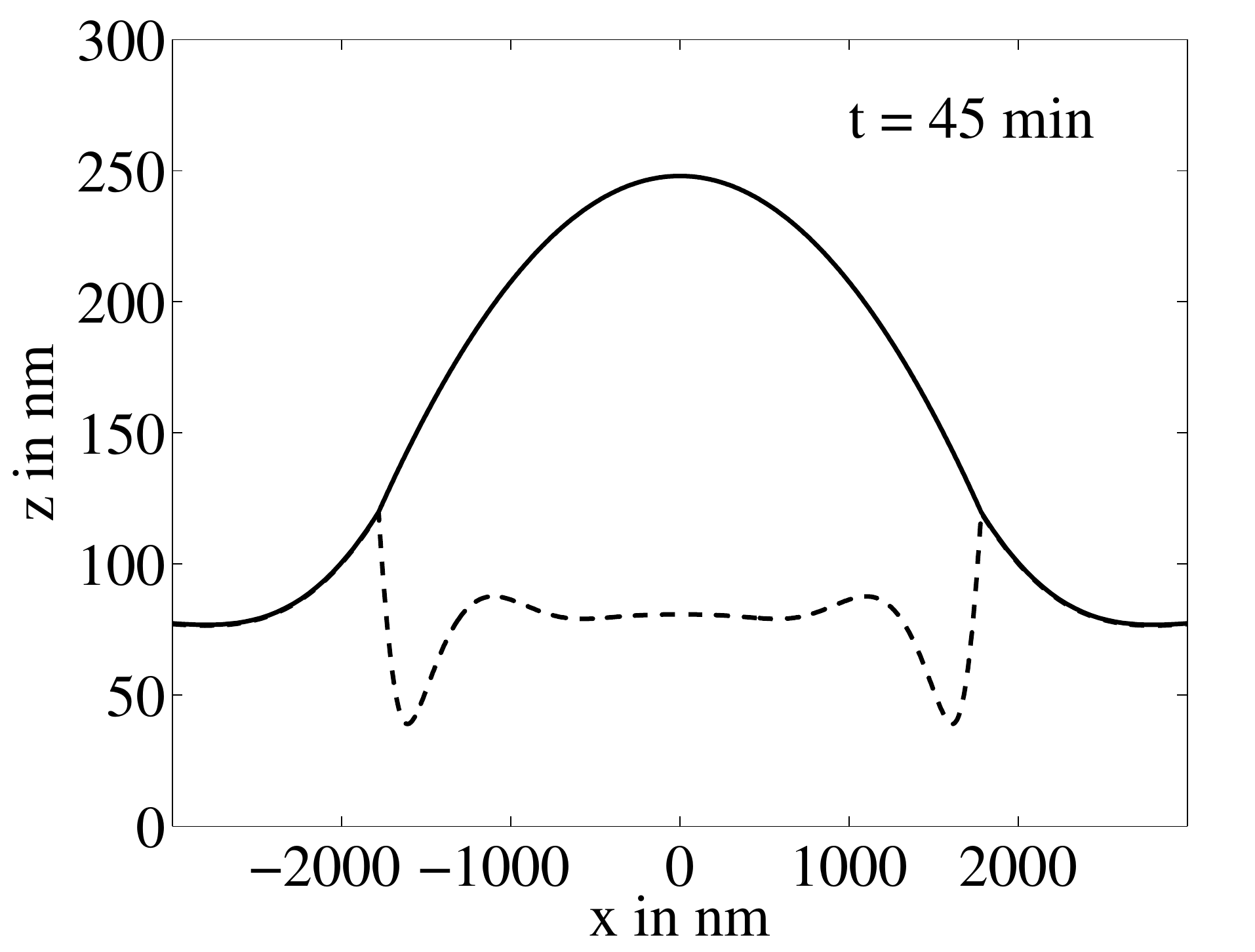}%
\includegraphics[width=0.24\textwidth]{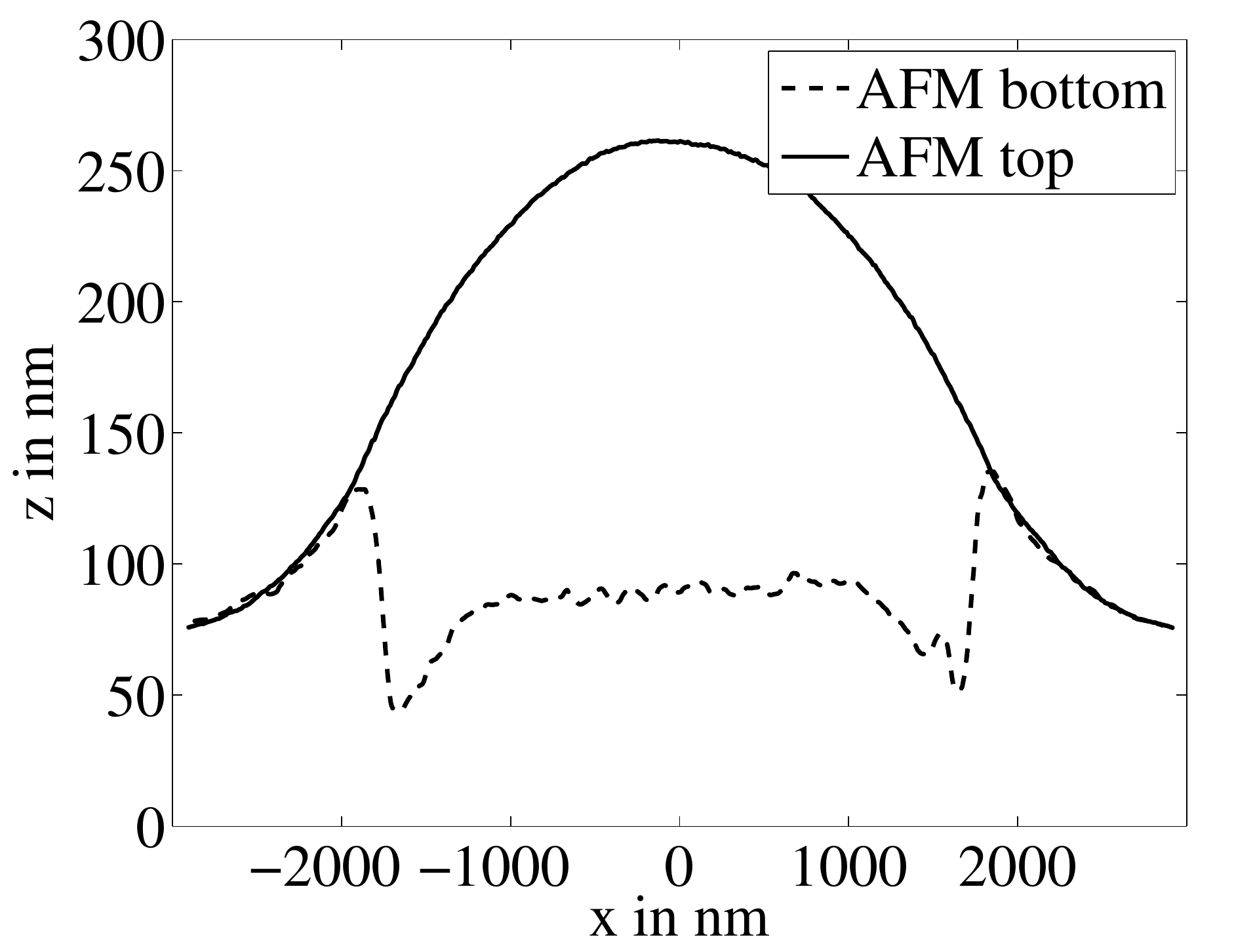}%

\includegraphics[width=0.24\textwidth]{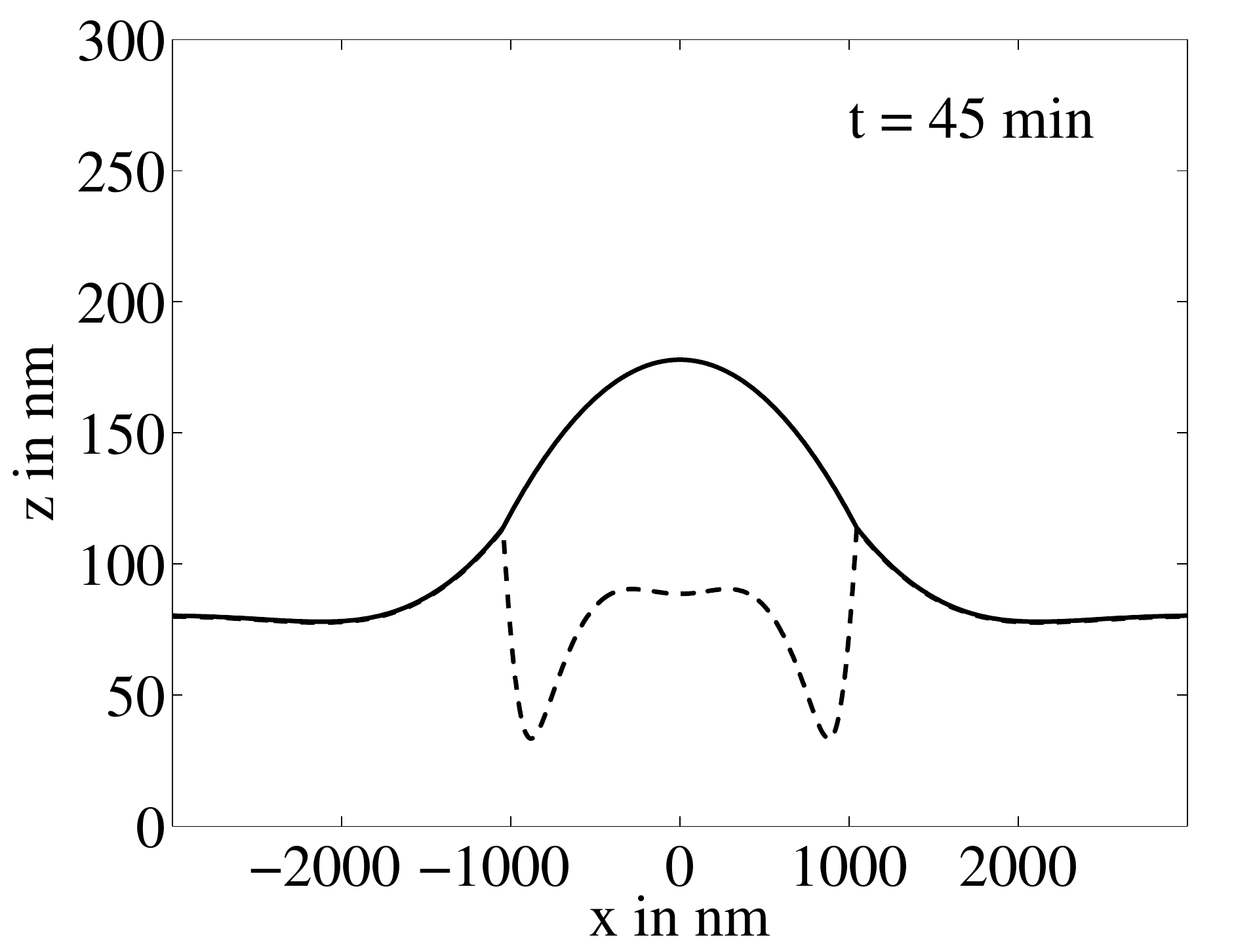}%
\includegraphics[width=0.24\textwidth]{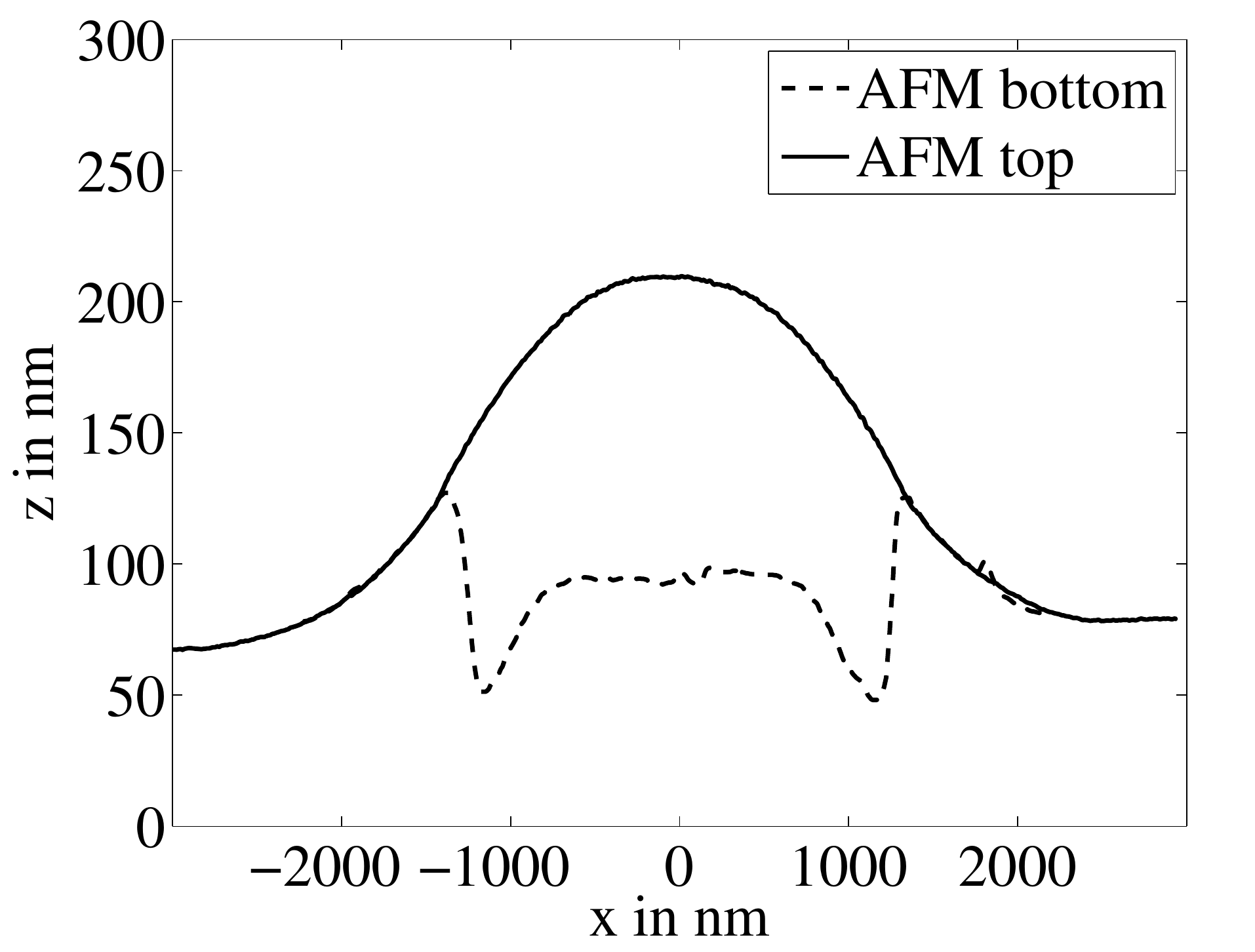}%

\includegraphics[width=0.24\textwidth]{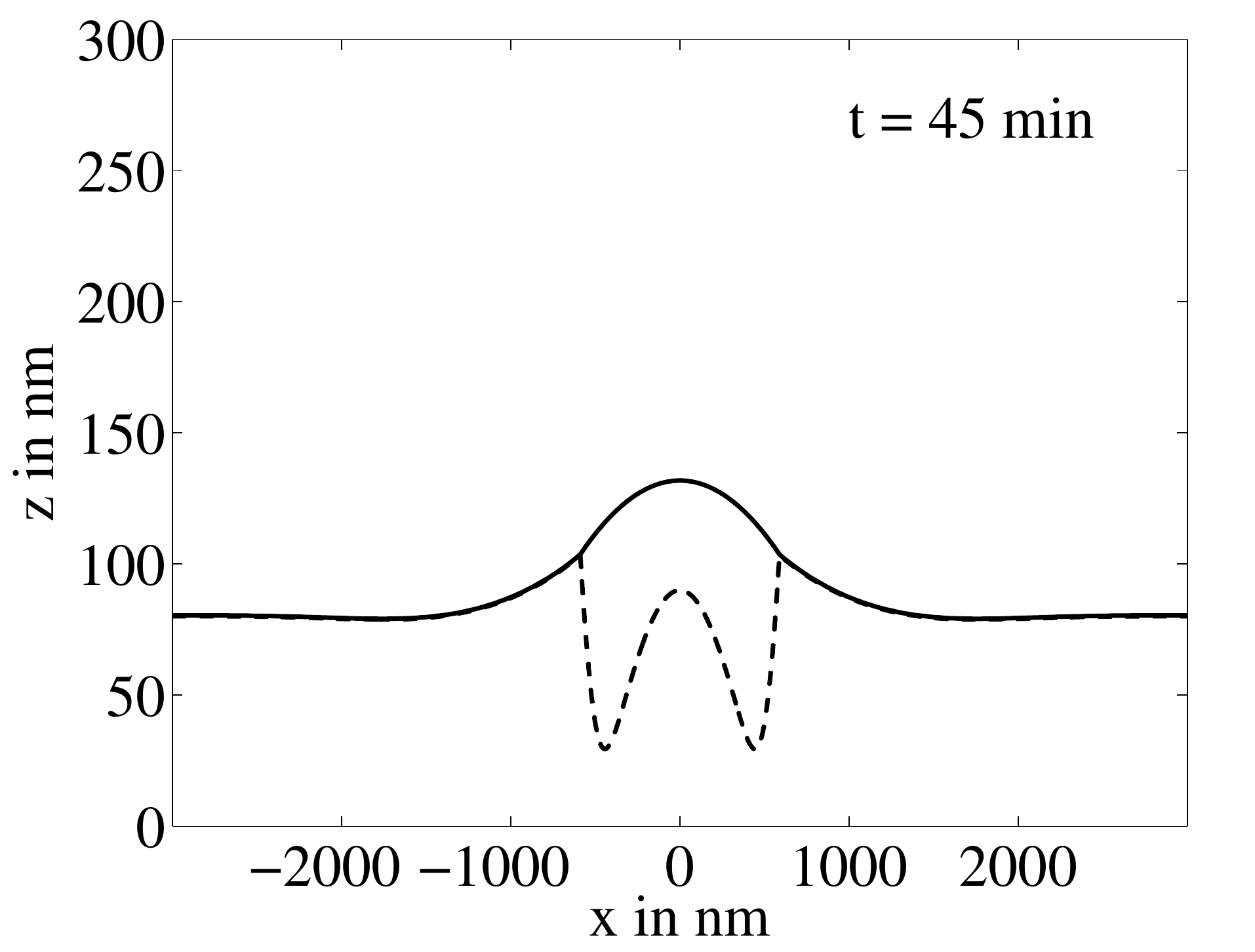}%
\includegraphics[width=0.24\textwidth]{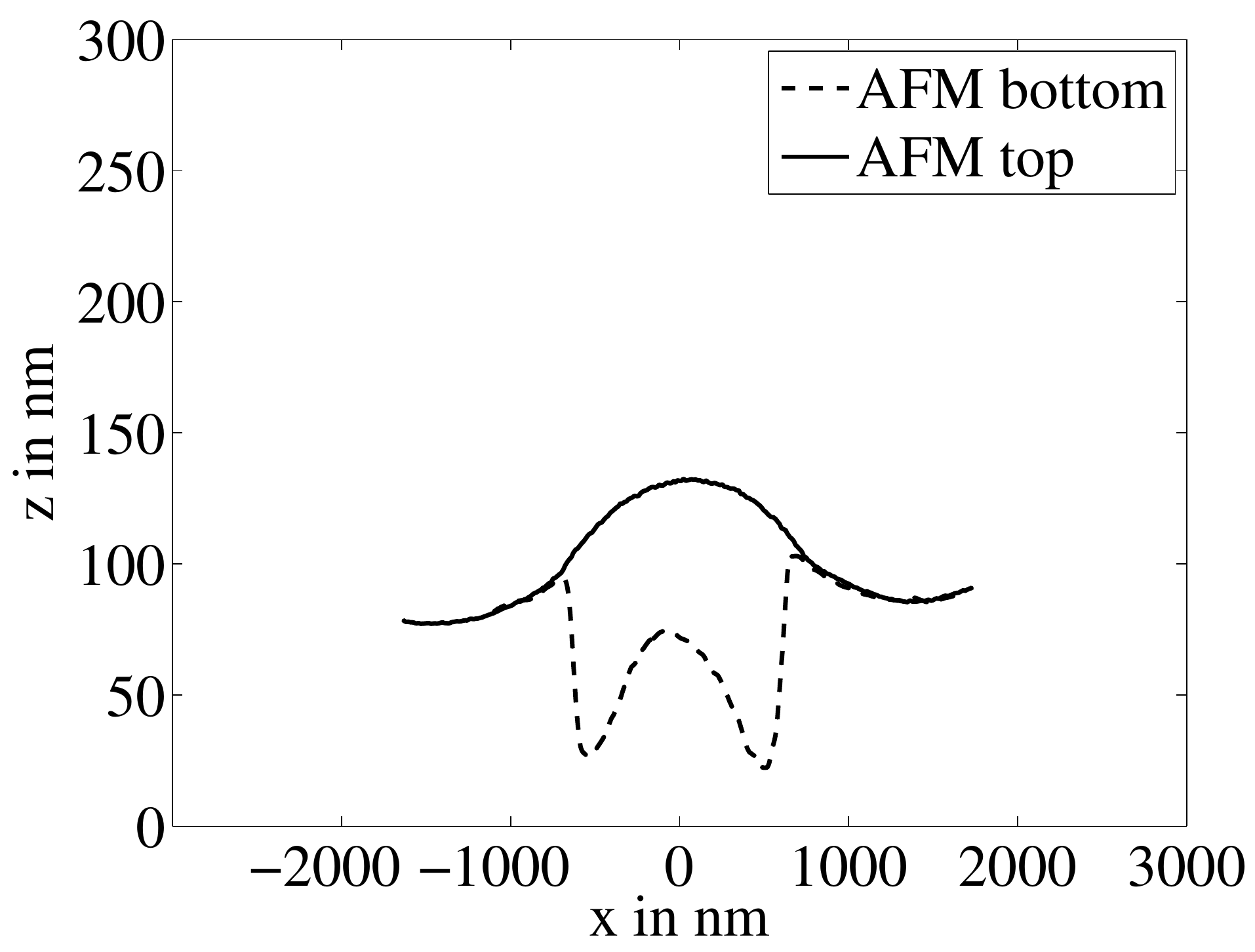}%

\includegraphics[width=0.24\textwidth]{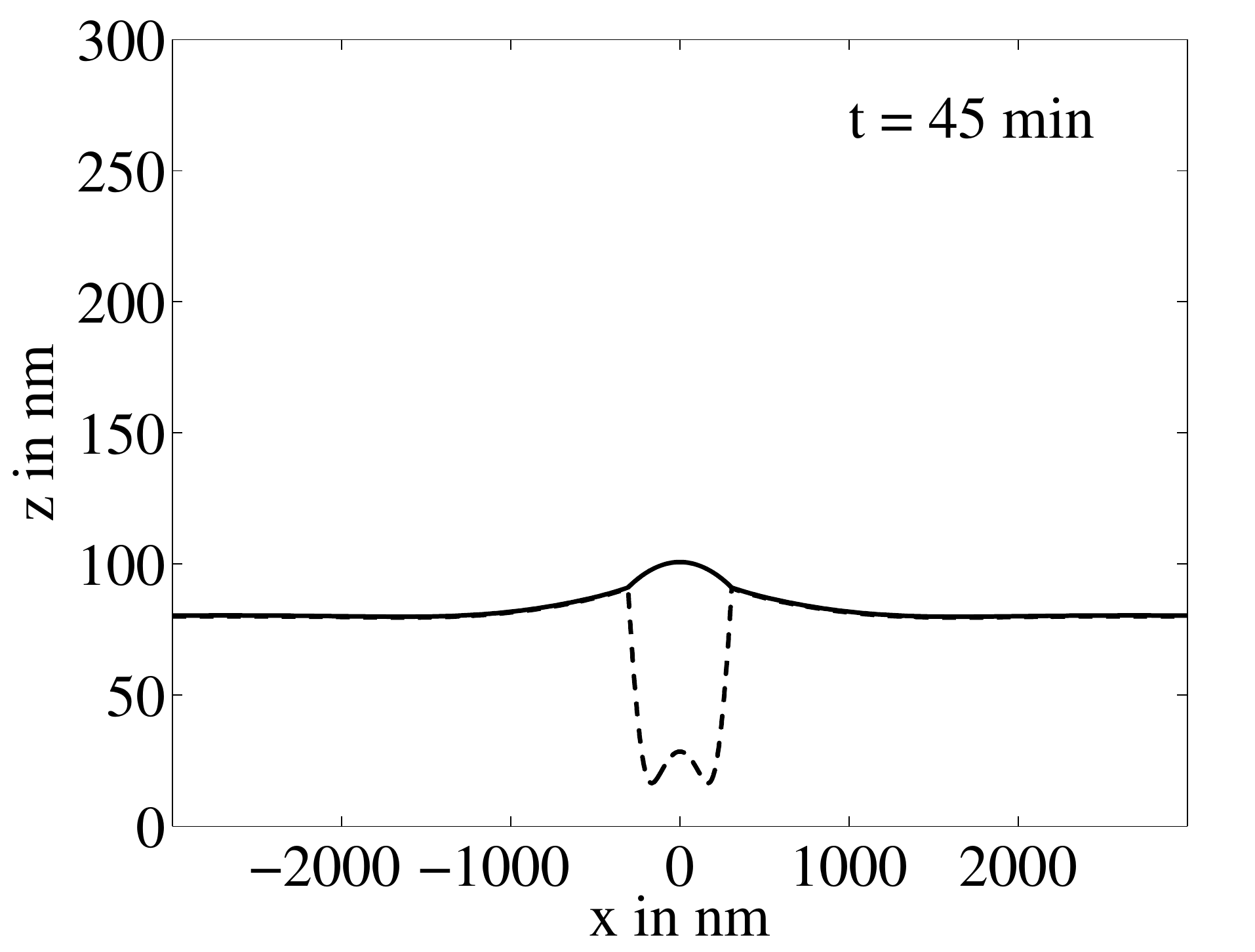}%
\includegraphics[width=0.24\textwidth]{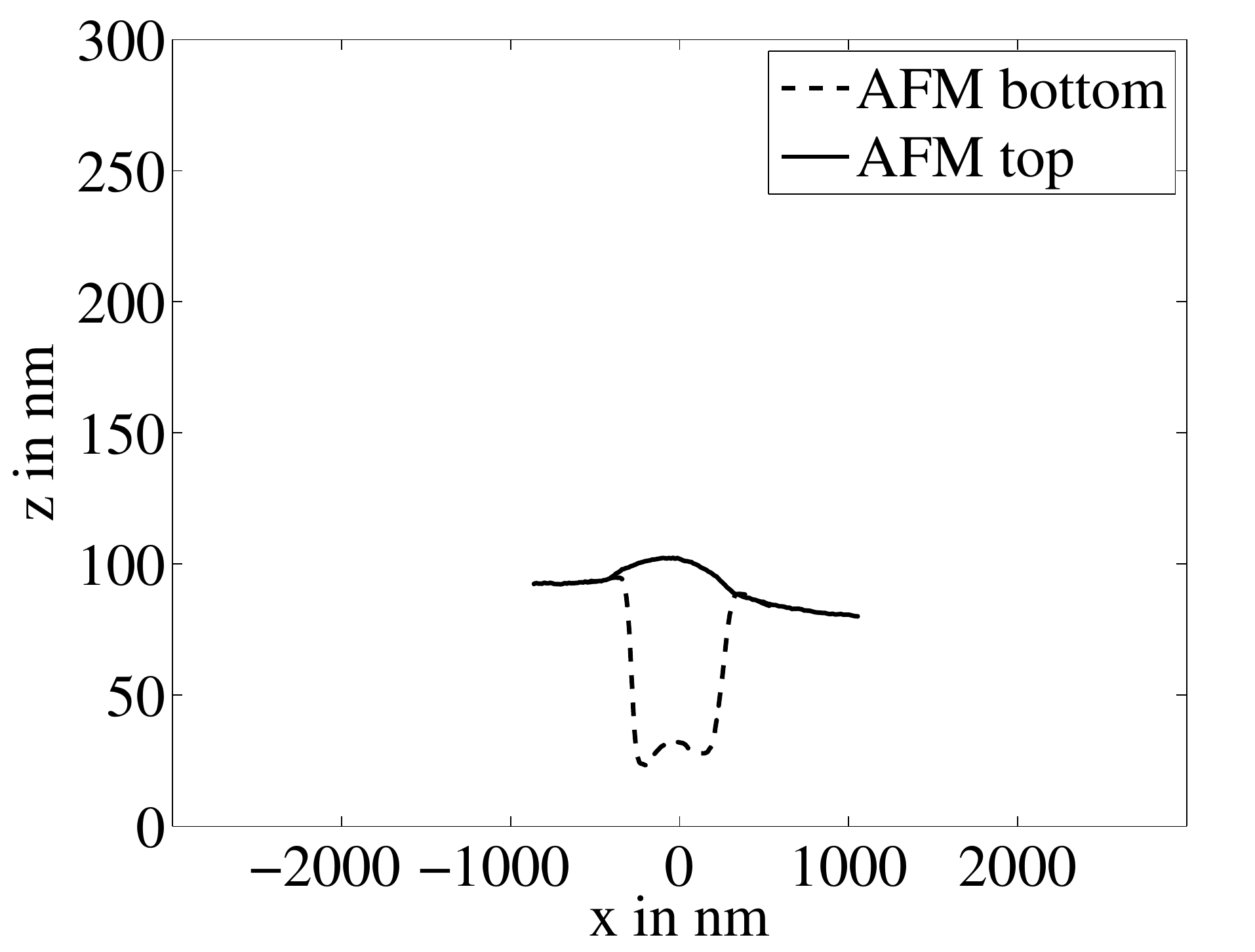}%

\caption{PS$(10k)$ droplets on PMMA$(10k)$ substrates at $T=140\celsius$ after dewetting a sample for $t=45\min$. The initially flat PMMA and PS layers had a thickness of $h_1=80\nm$ and $h_2-h_1=20\nm$, respectively. (\emph{left column}) simulations and (\emph{right column}) AFM measurements of similar volumes.}
\label{fig:earlystage}
\end{figure}

\begin{figure}
\centering
\includegraphics[width=0.24\textwidth]{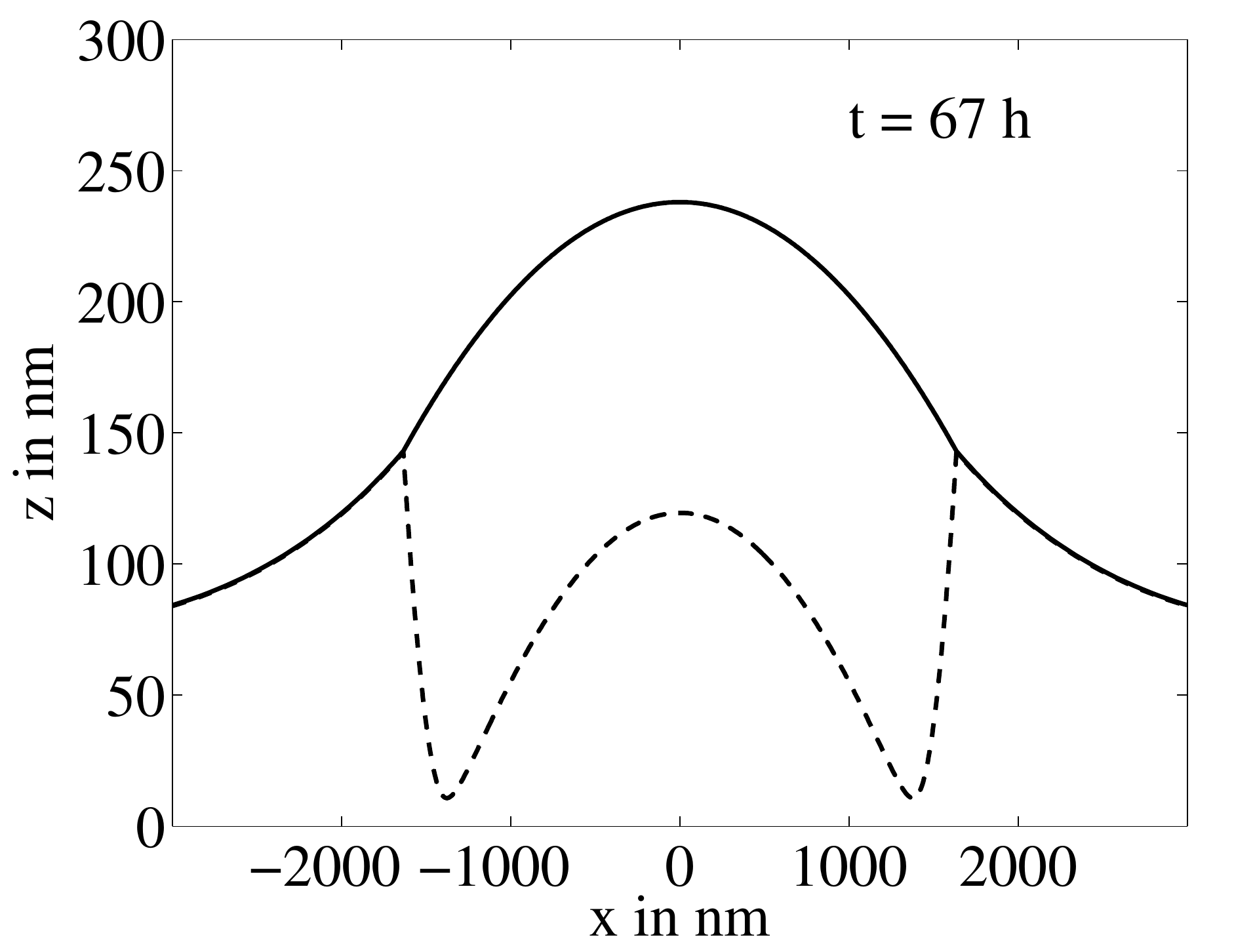}%
\includegraphics[width=0.24\textwidth]{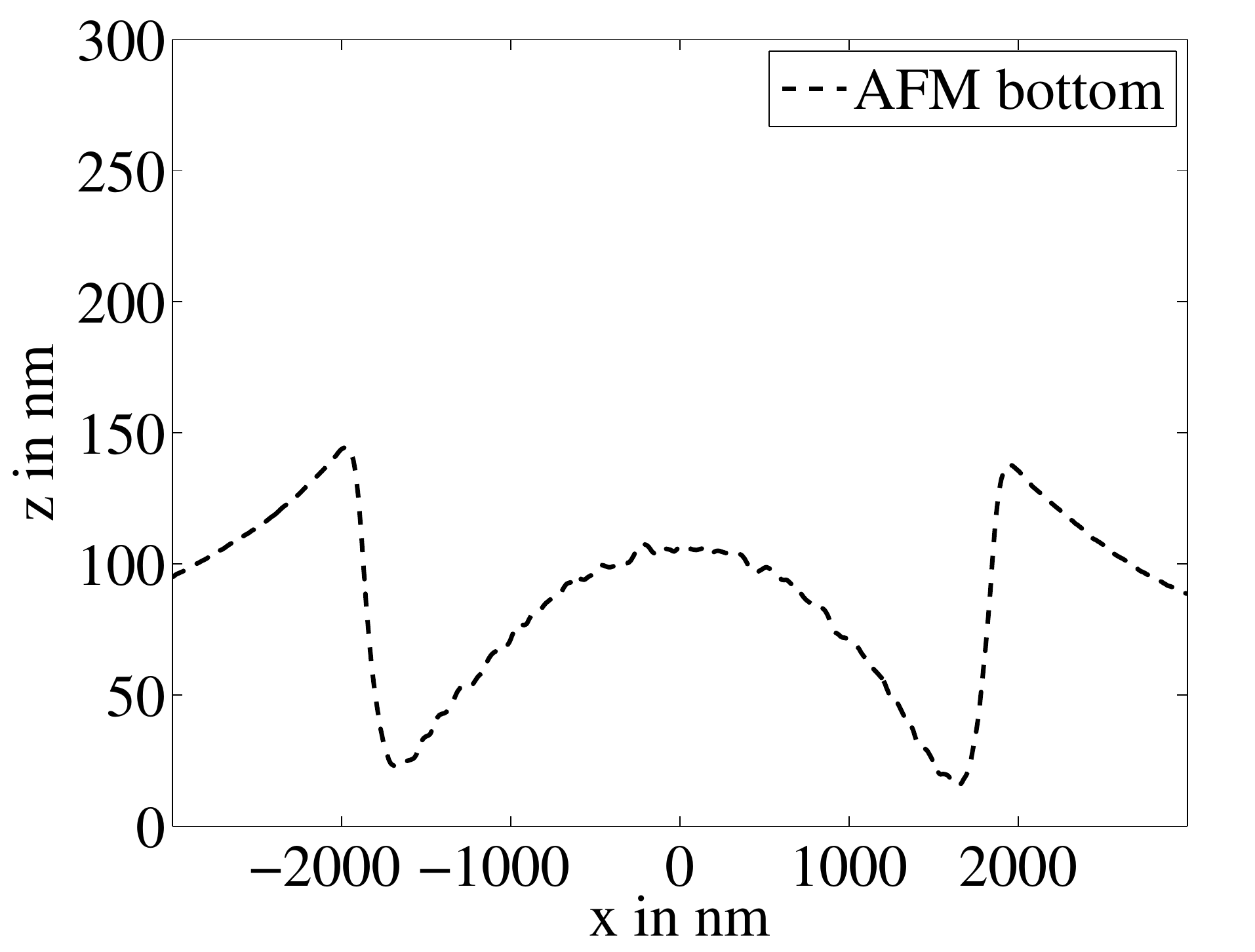}%

\includegraphics[width=0.24\textwidth]{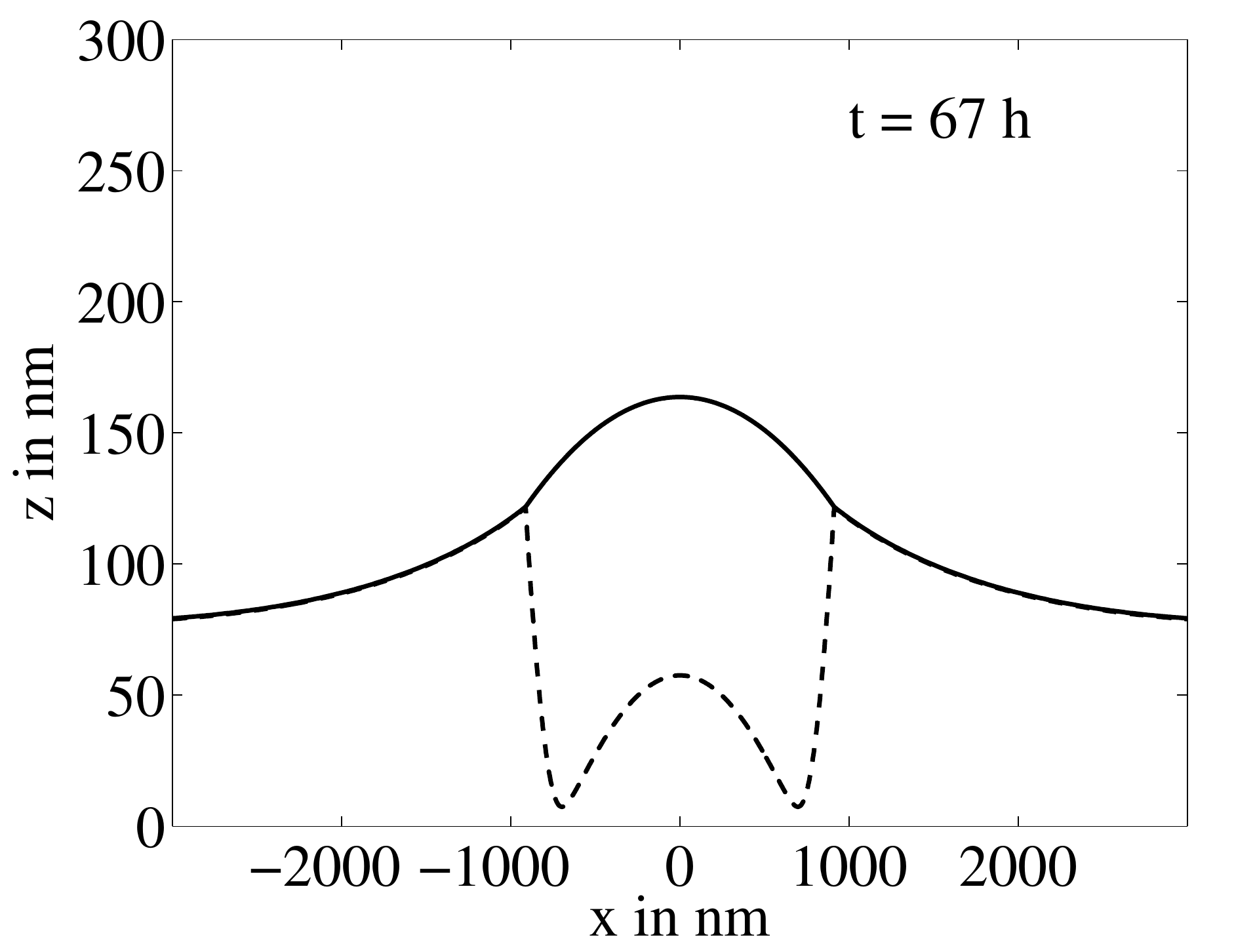}%
\includegraphics[width=0.24\textwidth]{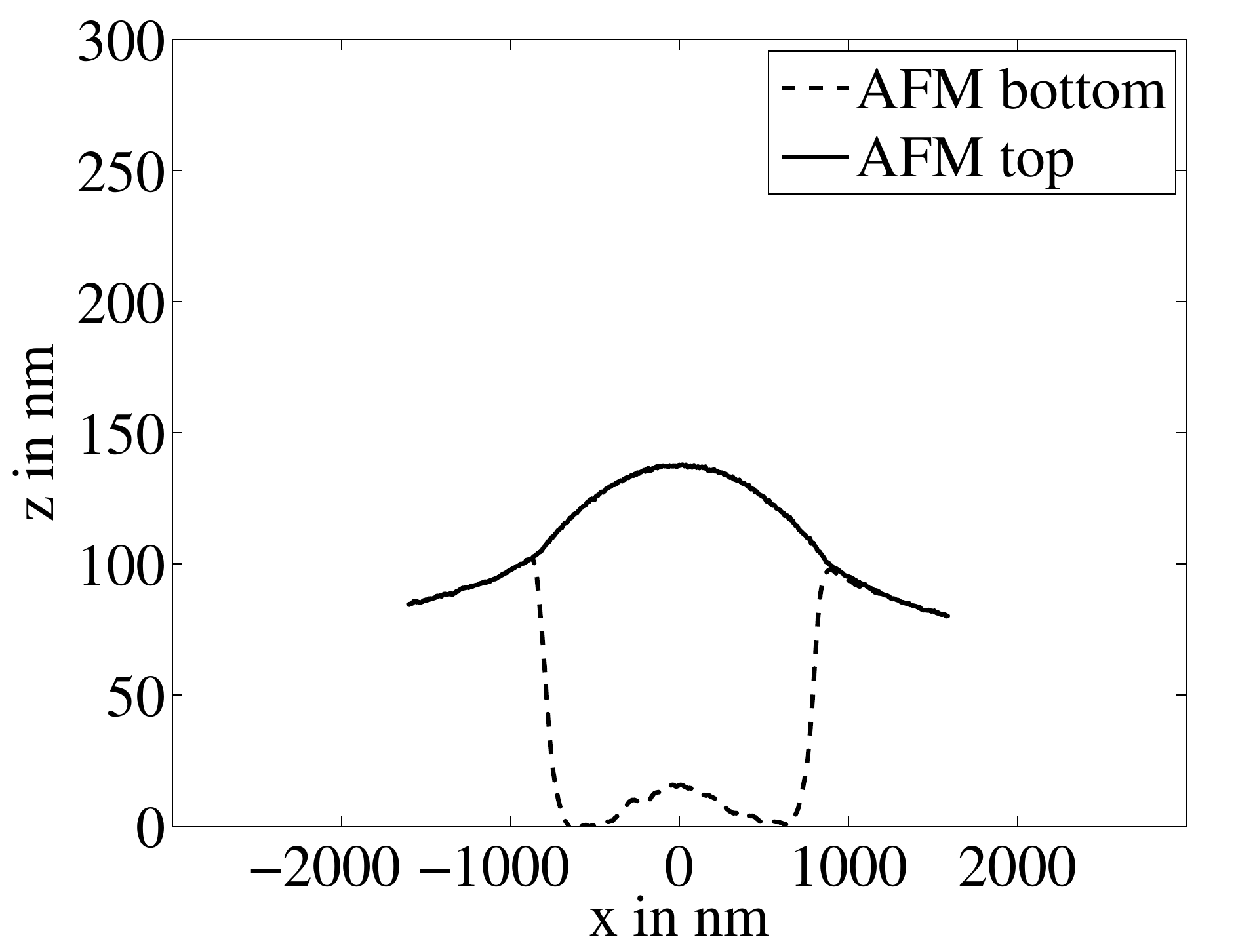}%

\includegraphics[width=0.24\textwidth]{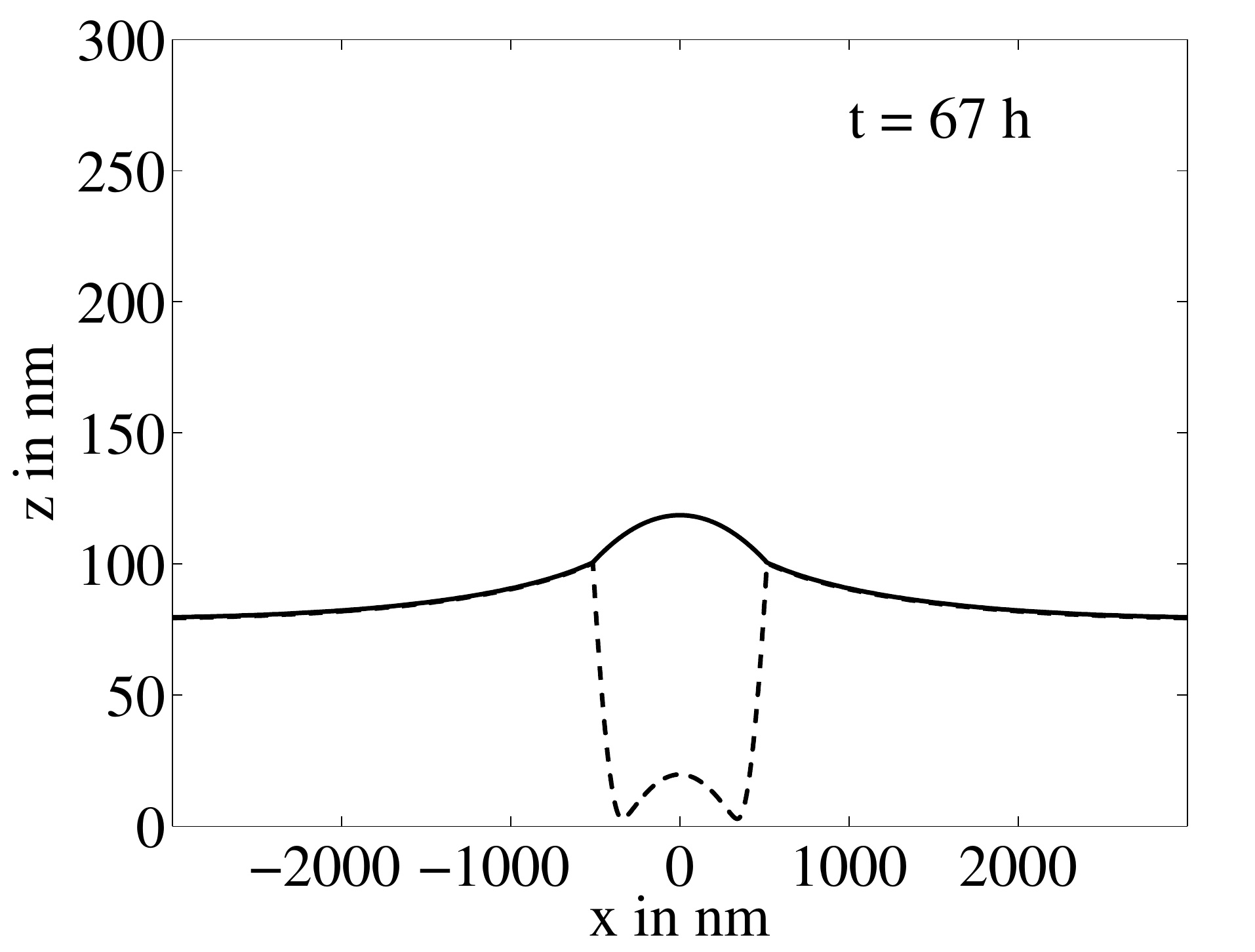}%
\includegraphics[width=0.24\textwidth]{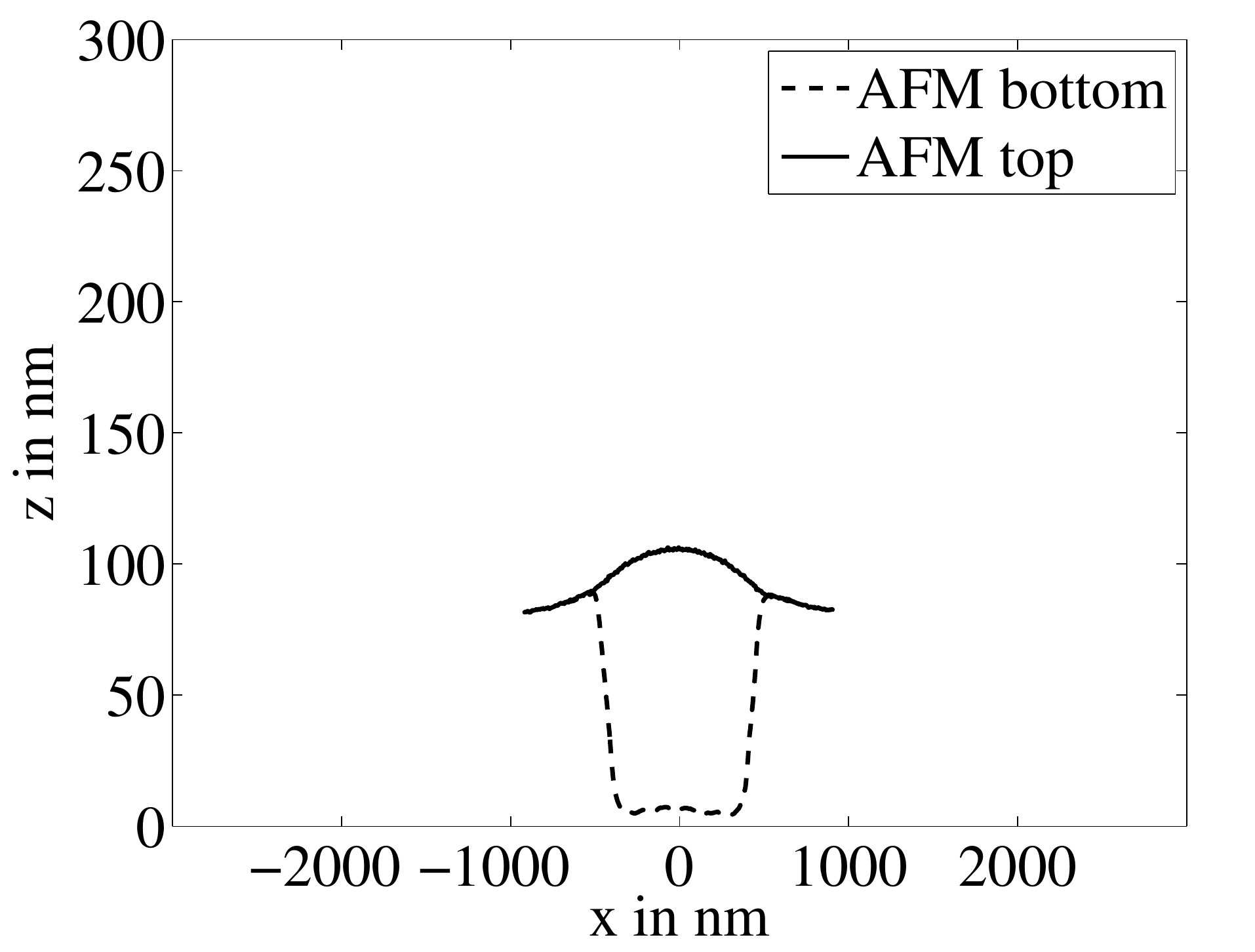}%

\includegraphics[width=0.24\textwidth]{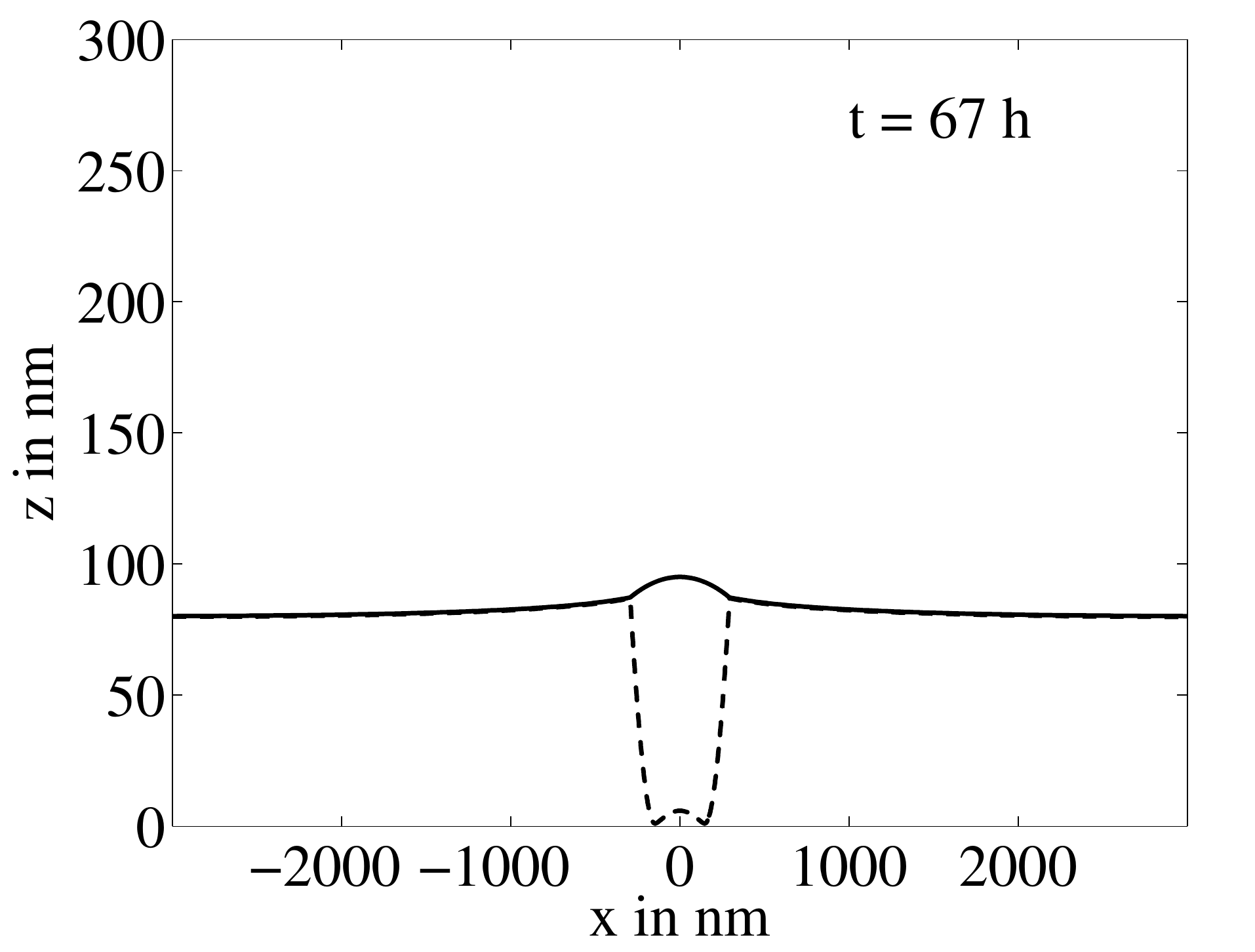}%
\includegraphics[width=0.24\textwidth]{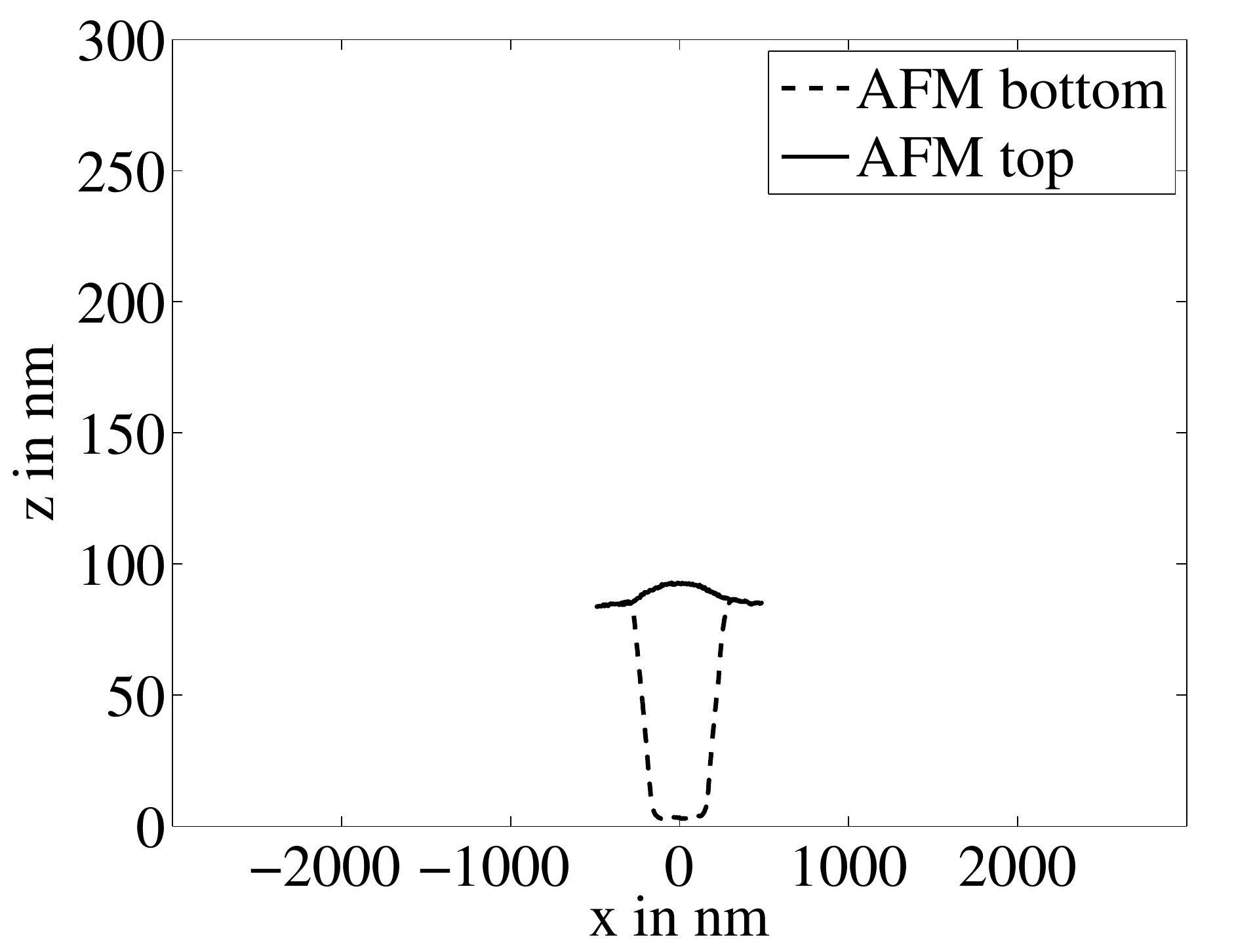}%

\includegraphics[width=0.24\textwidth]{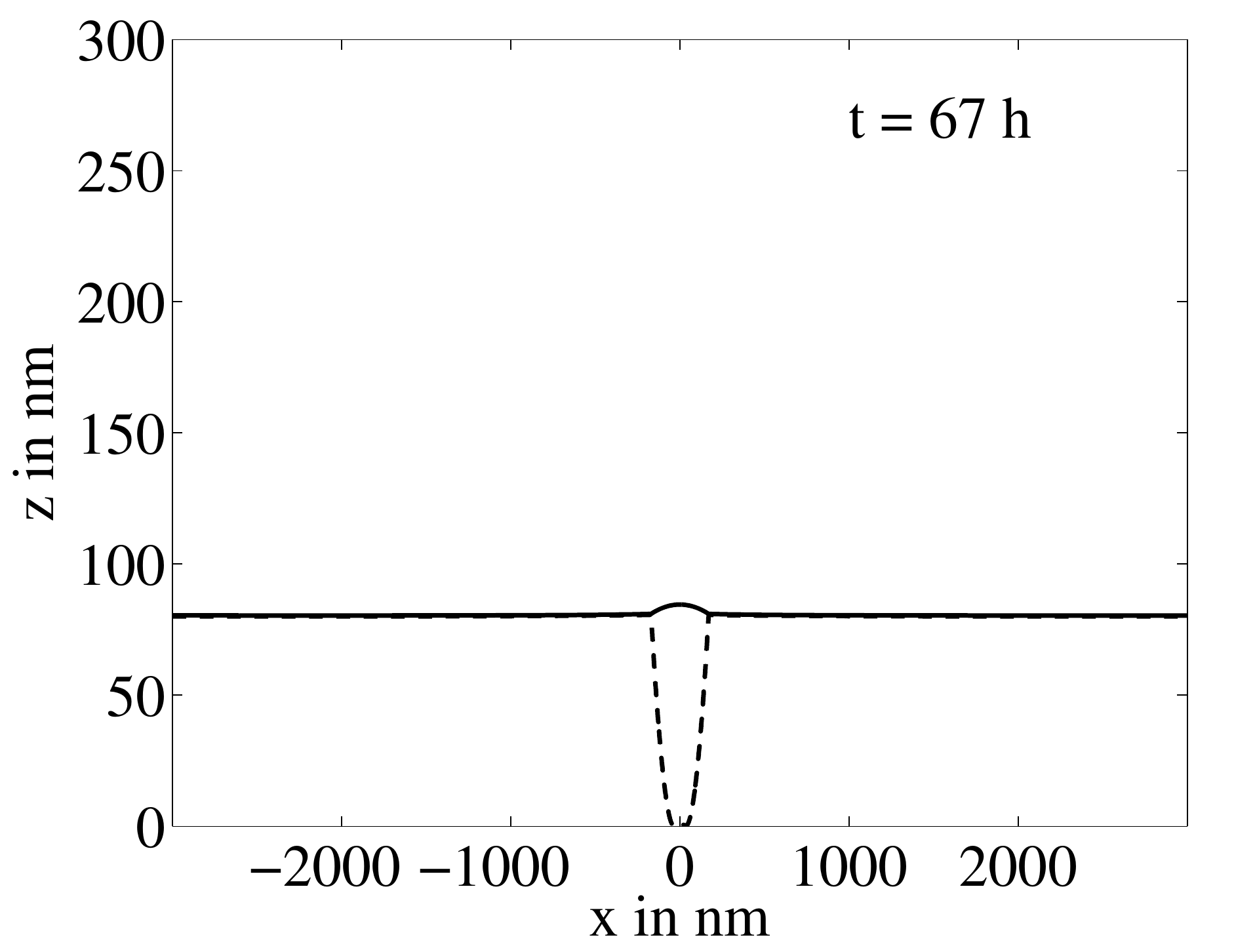}%
\includegraphics[width=0.24\textwidth]{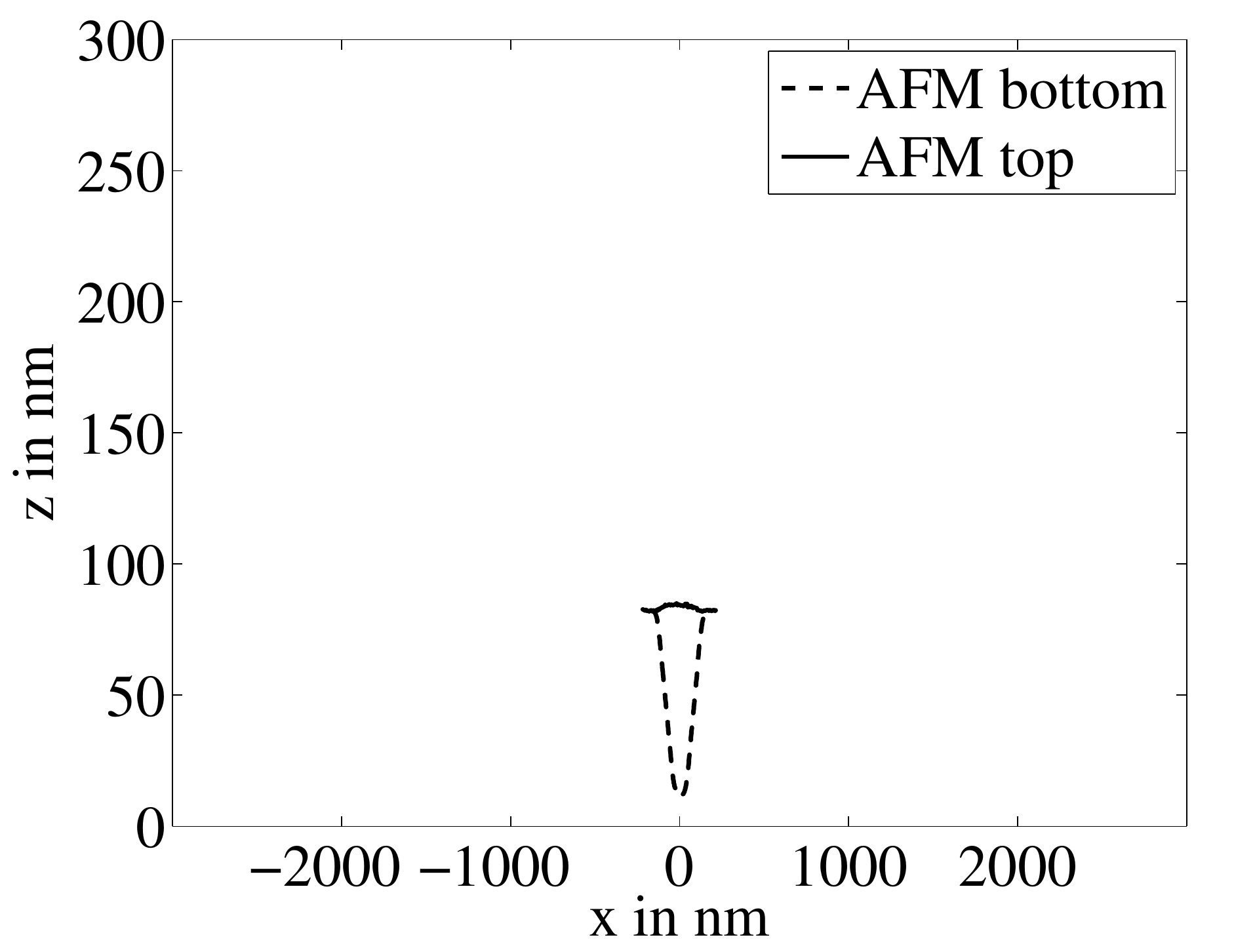}%

\caption{PS($10k$) droplets on PMMA($10k$) substrates at $T=140\celsius$ after dewetting a sample for $t=67\hour$. The initially flat PMMA and PS layers had a thickness of $h_1=80\nm$ and $h_2-h_1=20\nm$, respectively. (\emph{left column}) simulations and (\emph{right column}) AFM measurements of similar volumes.}
\label{fig:latestage}
\end{figure}

Figure~\ref{fig:earlystage} shows this comparison just just after having reached the time where the drop shapes are become independent from the initial  configuration. The experimental data compare quite well with the numerical solution of our lubrication model for all examined volumes verifying the extracted knowledge from the equilibrium droplets. Also for the late stages of the experiment shown in fig.~\ref{fig:latestage} this proves to be true. 

The remaining differences between simulation and experiment for larger times are suspected due be caused by possible slip effects at the substrate/PMMA interface, or intermolecular forces between the substrate and PMMA, which we so far have neglected in our model. This will be subject of our future investigations.

\section{Conclusion}

In this paper we considered the slow evolution of liquid PS droplets on liquid PMMA substrate into equilibrium. We extracted surface tensions from equilibrium morphologies, which was necessary for two reasons: Firstly, the values found in literature depend on temperature and molecular weight, whereas the parameter regimes studied in literature do not overlap with ours, \textit{i.e.} $T=140\celsius$ and
$M_w \approx 10\kg/\mole$ for PS and PMMA. Secondly, the sign of the spreading coefficient depends sensitively on the precise value of the surface tensions and is crucial for quantitative comparison.

The surface tension data obtained by this precise validation experiment is then used to compare experimentally measured surface/interface profiles with the ones obtained by numerically solving a thin-film equation for liquid two-layer films. Since the numerical solutions only depend on initial data through the total drop
volume, we performed quantitative comparisons and found very good agreement
for droplet shapes at similar times and drop volumes.

\section*{Acknowledgement}
All authors with to thank the DFG for financial support through the project
\textit{Structure formation in thin liquid-liquid films} in the SPP 1506.

\bibliographystyle{unsrtnat}
\bibliography{droplets}

\begin{thebibliography}{29}
\providecommand{\natexlab}[1]{#1}
\providecommand{\url}[1]{\texttt{#1}}
\expandafter\ifx\csname urlstyle\endcsname\relax
  \providecommand{\doi}[1]{doi: #1}\else
  \providecommand{\doi}{doi: \begingroup \urlstyle{rm}\Url}\fi

\bibitem[Zhang et~al.(2003)Zhang, Matar, and Craster]{ZMC03}
Y.~L. Zhang, O.K. Matar, and R.V. Craster.
\newblock {Analysis of tearfilm rupture: effect of non-Newtonian rheology}.
\newblock \emph{J. Coll. Int. Sci.}, 262:\penalty0 130--148, 2003.

\bibitem[Zhang et~al.(2005)Zhang, Matar, and Craster]{HJ05}
Y.~L. Zhang, O.K. Matar, and R.V. Craster.
\newblock An interfacial instability in a transient wetting layer leads to
  lateral phase separation in thin spin-cast polymer-blend films.
\newblock \emph{Nature Materials}, 4:\penalty0 782--786, 2005.

\bibitem[Lambooy et~al.(1996)Lambooy, Phelan, Haugg, and
  Krausch]{lambooy1996dewetting}
P.~Lambooy, K.C. Phelan, O.~Haugg, and G.~Krausch.
\newblock Dewetting at the liquid-liquid interface.
\newblock \emph{Physical Review Letters}, 76\penalty0 (7):\penalty0 1110--1113,
  1996.

\bibitem[Wang et~al.(2001)Wang, Krausch, and Geoghegan]{wang2001dewetting}
C.~Wang, G.~Krausch, and M.~Geoghegan.
\newblock Dewetting at a polymer-polymer interface: film thickness dependence.
\newblock \emph{Langmuir}, 17\penalty0 (20):\penalty0 6269--6274, 2001.

\bibitem[Brochard~Wyart et~al.(1993)Brochard~Wyart, Martin, and
  Redon]{brochard1993liquid}
F.~Brochard~Wyart, P.~Martin, and C.~Redon.
\newblock {Liquid/liquid dewetting}.
\newblock \emph{Langmuir}, 9\penalty0 (12):\penalty0 3682--3690, 1993.

\bibitem[Segalman and Green(1999)]{segalman1999dynamics}
R.A. Segalman and P.F. Green.
\newblock {Dynamics of rims and the onset of spinodal dewetting at
  liquid/liquid interfaces}.
\newblock \emph{Macromolecules}, 32\penalty0 (3):\penalty0 801--807, 1999.

\bibitem[Slep et~al.(2000)Slep, Asselta, Rafailovich, Sokolov, Winesett, Smith,
  Ade, and Anders]{slep00}
D.~Slep, J.~Asselta, M.~H. Rafailovich, J.~Sokolov, D.A. Winesett, A.P. Smith,
  H.~Ade, and S.~Anders.
\newblock {Effect of an Interactive Surface on the Equilibrium Contact Angles
  in Bilayer Polymer Films}.
\newblock \emph{Langmuir}, 16:\penalty0 2369--2375, 2000.

\bibitem[Pan et~al.(1997)Pan, Winey, Hu, and Composto]{pan1997unstable}
Q.~Pan, K.I. Winey, H.H. Hu, and R.J. Composto.
\newblock Unstable polymer bilayers. 2. the effect of film thickness.
\newblock \emph{Langmuir}, 13\penalty0 (6):\penalty0 1758--1766, 1997.

\bibitem[Li et~al.(2005)Li, Yang, Yu, and Dong]{li2005surface}
Y.~Li, Y.~Yang, F.~Yu, and L.~Dong.
\newblock Surface and interface morphology of polystyrene/poly (methyl
  methacrylate) thin-film blends and bilayers.
\newblock \emph{Journal of Polymer Science Part B: Polymer Physics},
  44\penalty0 (1):\penalty0 9--21, 2005.

\bibitem[Neto(2006)]{neto2006novel}
C.~Neto.
\newblock A novel approach to the micropatterning of proteins using dewetting
  of polymer bilayers.
\newblock \emph{Phys. Chem. Chem. Phys.}, 9\penalty0 (1):\penalty0 149--155,
  2006.

\bibitem[Higginsa et~al.(2002)Higginsa, Sferrazza, Jones, Jukes, Sharp, Dryden,
  and Webster]{HSJJSDW02}
A.~M. Higginsa, M.~Sferrazza, R.~A.~L. Jones, P.C. Jukes, J.S. Sharp, L.~E.
  Dryden, and J.~Webster.
\newblock {The timescale of spinodal dewetting at a polymer/polymer interface}.
\newblock \emph{Eur. Phys. J. E}, 8:\penalty0 137--143, 2002.

\bibitem[de~Silva et~al.(2007)de~Silva, Geoghegan, Higgins, Krausch, David, and
  Reiter]{de2007switching}
J.P. de~Silva, M.~Geoghegan, A.M. Higgins, G.~Krausch, M.O. David, and
  G.~Reiter.
\newblock {Switching layer stability in a polymer bilayer by thickness
  variation}.
\newblock \emph{PRL}, 98\penalty0 (26):\penalty0 267802, 2007.

\bibitem[Pototsky et~al.(2005)Pototsky, Bestehorn, Merkt, and
  Thiele]{pototsky2005morphology}
A.~Pototsky, M.~Bestehorn, D.~Merkt, and U.~Thiele.
\newblock Morphology changes in the evolution of liquid two-layer films.
\newblock \emph{The Journal of Chemical Physics}, 122\penalty0 (22):\penalty0
  224711, 2005.

\bibitem[Fisher and Golovin(2005)]{fisher2005nonlinear}
L.S. Fisher and A.A. Golovin.
\newblock {Nonlinear stability analysis of a two-layer thin liquid film:
  Dewetting and autophobic behavior}.
\newblock \emph{Journal of colloid and interface science}, 291\penalty0
  (2):\penalty0 515--528, 2005.

\bibitem[Bandyopadhyay et~al.(2005)Bandyopadhyay, Gulabani, and
  Sharma]{bandyopadhyay2005instability}
D.~Bandyopadhyay, R.~Gulabani, and A.~Sharma.
\newblock {Instability and dynamics of thin liquid bilayers}.
\newblock \emph{Ind. Eng. Chem. Res}, 44\penalty0 (5):\penalty0 1259--1272,
  2005.

\bibitem[Fisher and Golovin(2007)]{fisher2007instability}
L.S. Fisher and A.A. Golovin.
\newblock {Instability of a two-layer thin liquid film with surfactants:
  Dewetting waves}.
\newblock \emph{Journal of colloid and interface science}, 307\penalty0
  (1):\penalty0 203--214, 2007.

\bibitem[Seemann et~al.(2001{\natexlab{a}})Seemann, Herminghaus, and
  Jacobs]{seemann2001shape}
R.~Seemann, S.~Herminghaus, and K.~Jacobs.
\newblock Shape of a liquid front upon dewetting.
\newblock \emph{Physical Review Letters}, 87\penalty0 (19):\penalty0 196101,
  2001{\natexlab{a}}.

\bibitem[Becker et~al.(2003)Becker, Gr{\"u}n, Seemann, Mantz, Jacobs, Mecke,
  and Blossey]{becker2003complex}
J.~Becker, G.~Gr{\"u}n, R.~Seemann, H.~Mantz, K.~Jacobs, K.R. Mecke, and
  R.~Blossey.
\newblock Complex dewetting scenarios captured by thin-film models.
\newblock \emph{Nature Materials}, 2\penalty0 (1):\penalty0 59--63, 2003.

\bibitem[Seemann et~al.(2001{\natexlab{b}})Seemann, Herminghaus, and
  Jacobs]{seemann2001dewetting}
R.~Seemann, S.~Herminghaus, and K.~Jacobs.
\newblock Dewetting patterns and molecular forces: A reconciliation.
\newblock \emph{Physical Review Letters}, 86\penalty0 (24):\penalty0
  5534--5537, 2001{\natexlab{b}}.

\bibitem[Pechhold et~al.(1990)Pechhold, Grassl, and van Soden]{pechhold}
W.~Pechhold, O.~Grassl, and W.~van Soden.
\newblock Dynamic shear compliance of polymer meltsand networks.
\newblock In O.~G{\"u}ven, editor, \emph{Crosslinking and scission in
  polymers}, volume 292. Springer, 1990.

\bibitem[Fetzer et~al.(2007)Fetzer, M{\"u}nch, Wagner, Rauscher, and
  Jacobs]{fet2007}
R.~Fetzer, A.~M{\"u}nch, B.~Wagner, M.~Rauscher, and K.~Jacobs.
\newblock Quantifying hydrodynamic slip: A comprehensive analysis of dewetting
  profiles.
\newblock \emph{Langmuir}, 23\penalty0 (21):\penalty0 10559--10566, 2007.

\bibitem[Herminghaus et~al.(2001)Herminghaus, Jacobs, and
  Seemann]{herminghaus2001glass}
S.~Herminghaus, K.~Jacobs, and R.~Seemann.
\newblock The glass transition of thin polymer films: some questions, and a
  possible answer.
\newblock \emph{The European Physical Journal E: Soft Matter and Biological
  Physics}, 5\penalty0 (5):\penalty0 531--538, 2001.

\bibitem[B{\"a}umchen et~al.(2012)B{\"a}umchen, Fetzer, Klos, Lessel, Marquant,
  H{\"a}hl, and Jacobs]{baumchen2012slippage}
O.~B{\"a}umchen, R.~Fetzer, M.~Klos, M.~Lessel, L.~Marquant, H.~H{\"a}hl, and
  K.~Jacobs.
\newblock Slippage and nanorheology of thin liquid polymer films.
\newblock \emph{Journal of Physics: Condensed Matter}, 24\penalty0
  (32):\penalty0 325102, 2012.

\bibitem[van Krevelen(1976)]{krevelen1976}
D.W. van Krevelen.
\newblock \emph{Properties of Polymers: Their Estimation and Correlation with
  Chem. Structure}.
\newblock Elsevier Scientific Publ., 1976.

\bibitem[Anastasiadis et~al.(1988)Anastasiadis, Gancarz, and
  Koberstein]{anastasiadis1988interfacial}
S.H. Anastasiadis, I.~Gancarz, and J.T. Koberstein.
\newblock Interfacial tension of immiscible polymer blends: temperature and
  molecular weight dependence.
\newblock \emph{Macromolecules}, 21\penalty0 (10):\penalty0 2980--2987, 1988.

\bibitem[Wu(1970)]{wu1970surface}
S.~Wu.
\newblock {Surface and interfacial tensions of polymer melts. II. Poly (methyl
  methacrylate), poly (n-butyl methacrylate), and polystyrene}.
\newblock \emph{The Journal of Physical Chemistry}, 74\penalty0 (3):\penalty0
  632--638, 1970.

\bibitem[Jachalski et~al.(2012)Jachalski, Huth, Kitavtsev, Peschka, and
  Wagner]{jachalski2012stationary}
S.~Jachalski, R.~Huth, G.~Kitavtsev, D.~Peschka, and B.~Wagner.
\newblock Stationary solutions of liquid two-layer thin film models.
\newblock \emph{arXiv preprint arXiv:1210.5842}, 2012.

\bibitem[Kriegsmann and Miksis(2003)]{kriegsmann2003steady}
J.J. Kriegsmann and M.J. Miksis.
\newblock Steady motion of a drop along a liquid interface.
\newblock \emph{SIAM Journal on Applied Mathematics}, 64\penalty0 (1):\penalty0
  18--40, 2003.

\bibitem[M\"unch and Wagner(2005)]{MW05}
A.~M\"unch and B.~Wagner.
\newblock Contact-line instability of dewetting thin films.
\newblock \emph{Physica D}, 209:\penalty0 178--190, 2005.

\end{thebibliography}

\end{document}